\documentclass[prc,aps,superscriptaddress,twocolumn,showpacs,nofootinbib]{revtex4}

\usepackage{amsmath}
\usepackage{amssymb}
\usepackage{amsthm}
\usepackage{amsfonts}
\usepackage{graphicx}
\usepackage[usenames,dvipsnames]{color}
\usepackage{afterpage}
\usepackage{mathrsfs}
\usepackage{graphicx,latexsym,epsfig}
\usepackage{mathrsfs}
\usepackage{multirow}
\usepackage{color}
\usepackage{bm}
\usepackage[colorlinks,linkcolor=red,anchorcolor=blue,citecolor=blue]{hyperref}

\newcommand{\beq}{\vspace{0.5em}\begin{equation}}
\newcommand{\eeq}{\end{equation}\vspace{0.5em}}
\newcommand{\beqn}{\vspace{0.5em}\begin{eqnarray}}
\newcommand{\eeqn}{\end{eqnarray}\par\vspace{0.5em}\noindent}

\newcommand{\bsub}{\begin{subequations}}
\newcommand{\esub}{\end{subequations}}
\newcommand{\br}{\mathbf{r}}
\newcommand{\balp}{\bm{\alpha}}
\newcommand{\bp}{\mathbf{p}}
\newcommand{\bnab}{\bm{\nabla}}

\newcommand{\ttot}{\texttt{tot}}
\newcommand{\tovlp}{\texttt{overlap}}

\begin{document}

 \preprint{preprint}

\title{Beyond-mean-field study of the hyperon impurity effect in hypernuclei with shape coexistence}

 \author{X. Y. Wu} \email[]{xywu@stu.xmu.edu.cn}
\affiliation{Department of Physics and Institute of Theoretical Physics and Astrophysics, Xiamen University, Xiamen 361005, China}
 \author{H. Mei} \email[]{meihuayaoyugang@gmail.com}
\affiliation{School of Physical Science and Technology, Southwest University, Chongqing 400715, China}
 \author{J. M. Yao} \email[]{jmyao@unc.edu}
\affiliation{Department of Physics and Astronomy, University of North Carolina, Chapel Hill, NC 27516-3255, USA}
\affiliation{School of Physical Science and Technology, Southwest University, Chongqing 400715, China}
 \author{Xian-Rong Zhou} \email[]{xrzhou@phy.ecnu.edu.cn}
 \affiliation{Department of Physics, East China Normal University, Shanghai 200241, China}

 \date{\today}


 \begin{abstract}
\begin{description}
\item[Background] The hyperon impurity effect in nuclei has been extensively studied in different mean-field models. Recently, there is a controversy about whether the $\Lambda$ hyperon is more tightly bound in the normal deformed (ND) states than that in the superdeformed (SD) states.

\item[Purpose] This article is aimed to provide a beyond-mean-field study of the low-lying states of hypernuclei with shape coexistence and to shed some light on the controversy.

\item[Method] The models of relativistic mean-field and beyond based on a relativistic point-coupling energy functional are adopted to study the low-lying states of both $^{37}_\Lambda$Ar and $^{36}$Ar. The wavefunctions of low-lying states are constructed as a superposition of a set of relativistic mean-field states with different values of quadrupole deformation parameter. The projections onto both particle number and angular momentum are considered.

\item[Results] The $\Lambda$ binding energies in both ND and SD states of $^{37}_{\Lambda}$Ar are studied in the case of the $\Lambda$ hyperon occupying $s, p$, or $d$ state in the spherical limit, respectively. For comparison, four sets of nucleon-hyperon point-coupling interactions are used respectively. Moreover, the spectra of low-lying states in $^{36}$Ar and $^{37}_{\Lambda_s}$Ar are calculated based on the same nuclear energy density functional. The results indicate that the  SD states exist in $^{37}_{\Lambda}$Ar for all the four effective interactions. Furthermore, the $\Lambda_s$ reduces the quadrupole collectivity of ND states to a greater extent than that of SD states. For $^{37}_{\Lambda}$Ar, the beyond-mean-field decreases the $\Lambda_s$ binding energy of the SD state by 0.17 MeV, but it almost has no effect on that of the ND state.

\item[Conclusions]  In $^{37}_{\Lambda_s}$Ar, the $\Lambda_p$ and $\Lambda_d$ binding energies of the SD states are always larger than those of the ND states. For $\Lambda_s$, the conclusion depends on the effective nucleon-hyperon interaction. Moreover, the beyond-mean-field model calculation indicates that the $\Lambda_s$ hyperon is less bound in the SD state than that in the ND state.
\end{description}

\end{abstract}

\maketitle

%
 \section{Introduction}\label{Sec.I}
%
 The hyperon impurity effect in nuclear matter and atomic nuclei has attracted lots of attention since the first discovery of $\Lambda$ hypernuclei by Danysz and Pniewski in 1953~\cite{Danysz53,Danysz53-2}. A hyperon does not suffer from Pauli exclusion principle from nucleons and thus it can go deeply into the interior of nuclei and change remarkably nuclear properties (see, for example, Ref.~\cite{Hagino16} for a brief review). Previously, numerous studies have demonstrated that the presence of a $\Lambda$ hyperon may soften the equation of state of nuclear matter in neutron stars~\cite{Glendenning00} and changes nuclear structure significantly, such as nuclear shapes and sizes~\cite{Motoba83,Hiyama99,Tanida01,Win08,Win11}, collective excitations~\cite{Yao11,Isaka12,Mei14,Mei15,Xue15,Mei16}, neutron driplines~\cite{Vretenar98,Zhou08}, and fission barrier heights~\cite{Minato2009}.

  Shape coexistence exists universally in the nuclei throughout nuclear chart. For the nuclei around $A\sim40$ mass region, the coexistence of both ND and SD states was found in $^{36}$Ar~\cite{Svensson00,Svensson01}, $^{40}$Ca~\cite{Ideguchi01}, and $^{44}$Ti~\cite{OLeary00}, respectively. The structure of these states was studied in details theoretically~\cite{Caurier05,Caurier07,Inakura02,Bender03,Rodriguez04,Niksic06,Kimura04,Enyo05,Taniguchi09,Taniguchi10}.
  In recent years, the $\Lambda$ impurity effect on these nuclei has been studied in different mean-field based models. The extended antisymmetrized molecular dynamics model for hypernuclei (HyperAMD) predicted that the SD states exist in $^{41}_{\,\Lambda}$Ca and $^{46}_{\,\Lambda}$Sc~\cite{Isaka14}. In particular, the calculation indicates that the $\Lambda$ hyperon in the SD states is more bound than that in ND states. This study has generated a series of studies on hypernuclear SD states both in a non-relativistic framework~\cite{Isaka15,Zhou16} and a relativistic framework~\cite{Lu14}. However, whether the $\Lambda$ separation energy of the SD states is larger or smaller than that of the ND states is still a open question. For $^{37}_\Lambda$Ar, the HyperAMD model~\cite{Isaka15} and the Skyrme-Hartree-Fock (SHF) approach~\cite{Zhou16} predicted a smaller $\Lambda$ separation energy of the SD state, while the relativistic mean-field approach based on the meson-exchange (RMF-ME) effective nucleon-nucleon ($NN$) and nucleon-hyperon ($N \Lambda$) interactions with a finite-range separable pairing interaction~\cite{Tian09,Tian09-2,Niksic10} gave an opposite conclusion~\cite{Lu14}. According to Ref.~\cite{Lu14}, the larger $\Lambda$ binding energy in the SD state origin from a strong ring-shaped clustering structure which leads to a larger interaction energy between the nuclear core and the valence hyperon.

  Encouraged by the above discussion, we use the relativistic mean-field approach and beyond based on a point-coupling nucleon-nucleon and nucleon-hyperon effective interactions to study the effect of hyperon in $^{37}_\Lambda$Ar. The paper is organized as follows. In Sec.~\ref{Sec.II}, we briefly describe the point-coupling relativistic mean-field and beyond approach for single-$\Lambda$ hypernuclei. The numerical details are given in Sec.~\ref{Sec.III}. In Sec.~\ref{Sec.IV}, we present the results for the normal deformed (ND) and superdeformed (SD) states in $^{37}_\Lambda$Ar. Finally, a summary of our work is given in Sec.~\ref{Sec.V}.

%
 \section{Theoretical framework}
 \label{Sec.II}
%
%
 \subsection{Relativistic mean-field model}
 \label{Subsec.A}
 The mean-field states are obtained by the triaxially deformed relativistic mean-field model with point-coupling (RMF-PC) for $\Lambda$ hypernuclei. For details, please refer to Ref.~\cite{Xue15}. Here, we just present an outline of this model.

 The RMF-PC model for $\Lambda$ hypernuclei starts from an effective Lagrangian density
 \beqn
   \label{Lagrangian}
   {\cal L} ={\cal L}^{\rm free} + {\cal L}^{\rm em} + {\cal L}^{NN} + {\cal L}^{N\Lambda},
 \eeqn
 where the first term ${\cal L}^{\rm free}$ denotes the free Lagrangian density of hypernuclear system. The second term ${\cal L}^{\rm em}$ is an electromagnetic part for protons. The third term ${\cal L}^{NN}$ takes the standard form~\cite{Burvenich02} for the nucleon-nucleon effective interaction. The last term ${\cal L}^{N\Lambda}$ for nucleon-hyperon effective interaction is chosen as the form proposed in Ref.~\cite{Tanimura12}.

 From the Lagrangian density Eq.~(\ref{Lagrangian}), one obtains the corresponding energy density function $E_{\textrm{RMF}}$ at the mean-field level, which can be decomposed into two parts: the pure nucleonic part $E^{N}_{\textrm{RMF}}$ and the other part due to the presence of $\Lambda$ hyperon $E^{\Lambda}_{\textrm{RMF}}$,
 \begin{eqnarray}
 E^{N}_{\textrm{RMF}}  &=& T_N + \int d^3r\varepsilon_{NN}(\textbf{r})+ \frac{1}{2}A_0 e\rho_V^{(p)}, \\
 E^{\Lambda}_{\textrm{RMF}}
 &=& T_\Lambda +\int d^3r\varepsilon_{N\Lambda}(\textbf{r}),
 \end{eqnarray}
 where the first term $T_{B=N/\Lambda}={\rm Tr}[(\vec{\alpha}\cdot\vec{p}+m_B\beta)\rho^B_V]$ is for the kinetic energy of nucleons or $\Lambda$ hyperon.
 $A_0$ is for the time-like component of electromagnetic field and $\rho_V^{(p)}$ for the vector density of protons. The interaction energy terms are as follows
 \begin{eqnarray}\nonumber
 \label{Energy:Nh}
 \varepsilon_{NN}&=\!\!&\frac{1}{3}\beta_S(\rho_S^N)^3+\frac{1}{4}\gamma_S(\rho_S^N)^4 +\frac{1}{4}\gamma_V(\rho_V^N)^4\\
 &\!\!&+\frac{1}{2}\sum\limits_{K=S,V,TV}[\alpha_K(\rho_K^N)^2+ \delta_K\rho^N_K\Delta\rho_K^N],\\ \nonumber
 \label{Energy:NL}\varepsilon_{N\Lambda}&=\!\!&\sum\limits_{K=S,V}\alpha_K^{(N\Lambda)}\rho_K^N\rho_K^\Lambda+\sum\limits_{K=S,V}\delta_S^{(N\Lambda)}\rho_K^N\Delta\rho_K^\Lambda\\
  &\!\!&+\alpha_T^{(N\Lambda)}\rho_V^N\rho_T^\Lambda,
 \end{eqnarray}
 where the densities are defined as
 \begin{eqnarray}
 \rho^N_S&=&\sum \limits_k\bar\psi^N_k\psi^N_k , \hspace{1cm}
 \rho^N_V=\sum \limits_k\psi^{N\dagger}_k\psi^N_k ,\hspace{1cm}\\
 \rho^N_{TS}&=&\sum\limits_k\bar\psi^N_k\tau_3\psi^N_k ,\hspace{0.5cm}
 \rho^N_{TV}=\sum\limits_k\psi^{N\dagger}_k\tau_3\psi^N_k,\\
 \rho^\Lambda_S&=&\sum \limits_k\bar\psi^\Lambda_k\psi^\Lambda_k,\hspace{1cm}
 \rho^\Lambda_V=\sum \limits_k\psi^{^\Lambda\dagger}_k\psi^\Lambda_k ,\\
 \rho^\Lambda_T&=&\nabla\cdot(\bar\psi_\Lambda i\vec{\alpha}\psi_\Lambda).
 \end{eqnarray}

 The indices $S, V$, and $TV$ represent the symmetry of the coupling. The subscript $S$ stands for isoscalar-scalar, $V$ for isoscalar-vector, and $TV$ for isovector-vector type of coupling characterized by their transformation properties in isospin and in space-time.

 Minimization of the total energy with respect to the single-particle wavefunction $\psi^{B}_k(\br)$
 of nucleon or hyperon leads to Dirac equation,
 \beq
   \label{Dirac:NL}
   \left[\balp\cdot\bp+V^B_0 +\gamma^0(m_B+S^B)\right]\psi^B_k(\br)=\epsilon_k^B\psi^B_k(\br).
 \eeq
 For nucleons ($B=N$), the scalar field  $S^N(\br)=\Sigma_S(\br)+\tau_3 \Sigma_{TS}(\br)$ and the vector field
 $V^N_0(\br)=\Sigma_V(\br)+\tau_3\Sigma_{TV}(\br)$ take the standard form
 \bsub
  \beqn
   \label{dirac-nucleon1}
   \Sigma_S\!\!&=&\!\!\alpha_S\rho_S^N+\beta_S(\rho_S^N)^2+\gamma_S(\rho_S^N)^3+\delta_S\Delta\rho_S^N \nonumber \\
   &&\!\!+\alpha_S^{(N\Lambda)}\rho_S^\Lambda+\delta_S^{(N\Lambda)}\Delta\rho_S^\Lambda
   \, , \\
   \label{dirac-nucleon2}
   \Sigma_{TS}\!\!&=&\!\!\delta_{TS}\Delta\rho_{TS}^N+\alpha_{TS}\rho_{TS}^N
   \, , \\
   \label{dirac-nucleon3}
   \Sigma_V\!\!&=&\!\!\alpha_V\rho_V^N+\gamma_V(\rho_V^N)^3+\delta_V\Delta\rho_V^N
   +eA_0\frac{1-\tau_3}{2} \nonumber \\
   &&\!\!+\alpha_V^{(N\Lambda)}\rho_V^\Lambda+\delta_V^{(N\Lambda)}\Delta\rho_V^\Lambda+\alpha_T^{(N\Lambda)}\rho_T^\Lambda
   \, , \\
   \label{dirac-nucleon4}
   \Sigma_{TV}\!\!&=&\!\!\alpha_{TV}\rho_{TV}^N+\delta_{TV}\Delta\rho_{TV}^N
   \, .
  \eeqn
 \esub
 For $\Lambda$ hyperon ($B=\Lambda$), the scalar field $S^\Lambda(\br)$ and the vector field
 $V^\Lambda_0(\br)=U_V(\br)+ U_T(\br)$ are defined as
 \bsub
  \beqn
   \label{dirac-Lambda}
   S^\Lambda &=&\delta_S^{(N\Lambda)}\Delta\rho_S^N+\alpha_S^{(N\Lambda)}\rho_S^N
   \, ,\\
   \label{U_V}
   U_V&=&\delta_V^{(N\Lambda)}\Delta\rho_V^N+\alpha_V^{(N\Lambda)}\rho_V^N
   \, ,\\
   \label{U_T}
   U_T&=&-i\alpha_T^{(N\Lambda)}\beta\balp\cdot \bnab\rho_V^N
   \, .
  \eeqn
 \esub
 In Eq.~(\ref{Dirac:NL}), the $\epsilon_k^{B}$ is the single-particle energy of either nucleons or $\Lambda$ hyperon.

 A quadratic constraint calculation of the mass quadrupole moment $\langle {\hat q_{20}}\rangle=\sqrt{\frac{5}{16\pi}}\,\langle 2z^2-x^2-y^2 \rangle$ is carried out. The intrinsic deformation is defined as $\beta=\frac{4\pi}{3AR_0^2}\langle {\hat q_{20}}\rangle$ with $R_0=1.2\times A_c^{1/3}$ fm,
  and $A_c=A-1$ is the mass number of the core nucleus (cn).
 The deformation parameters $\beta$ are calculated either with the nuclear density $\rho^N(\br)$
 for the core nucleus or with the total density $\rho^N(\br)$+$\rho^\Lambda(\br)$
 for the hypernucleus.

%
 \subsection{Generator coordinate method with quantum number projections} \label{Subsec.B}
%
%
 The wavefunctions for the low-lying states of hypernuclei are constructed as the superpositions of a set of quadrupole deformed hypernucler mean-field states with particle number and angular momentum projection. This framework is known as PNAMP+GCM scheme and has been developed in Ref.~\cite{Mei16} for the hypernuclear systems composed of a $\Lambda$ hyperon and an even-even nuclear core. The wavefunction $|\Psi^{JM}_{n\alpha} \rangle$ reads
 \beq
 \label{GCM:wf}
 | \Psi^{JM}_{n\alpha} \rangle
 = \sum _{\beta} f^{J}_{n\alpha}(\beta)\hat{P}^J_{MK}\hat{P}^N \hat{P}^Z|\Phi^{(N\Lambda)}_{n}(\beta)\rangle,
 \eeq
 with $\hat{P}^J_{MK}$, $\hat{P}^N $, and $\hat{P}^Z$ being the angular momentum projection operators for neutrons and protons, respectively.
 The index $n$ refers to a different hyperon orbital state, and the index $\alpha$ labels the quantum numbers of the states other than the angular momentum.

 Since in hypernuclei the hyperon and nucleons are not mixed, the mean-field states $|\Phi^{(N\Lambda)}_{n}(\beta)\rangle$ can be decomposed into two parts
 \beq
 \label{Refst}
 |\Phi^{(N\Lambda)}_{n}(\beta)\rangle
 = |\Phi^N(\beta)\rangle \otimes |\varphi^{\Lambda}_{n}(\beta)\rangle,
 \eeq
 where $|\Phi^N(\beta)\rangle $ and $|\varphi^{\Lambda}_{n}(\beta)\rangle$ are the mean-field wavefunctions for nuclear core and the hyperon, respectively. They are Slater determinants built upon single-particle spinors $\psi^{B=N,\Lambda}_k(\br)$ from Eq.~(\ref{Dirac:NL}).

 The weight function $f^{J }_{n\alpha}(\beta)$ in the GCM states
 given by Eq.~(\ref{GCM:wf}) is determined by the variational principle which leads to the
 Hill-Wheeler-Griffin (HWG) equation,
  \beq
  \label{HWE}
  \sum_{\beta'}
  \left[{\cal H}^J_{n}(\beta,\beta') -E^{J}_{n\alpha}
 {\cal N}^J_{n }(\beta,\beta')\right]
   f^{J}_{n\alpha}(\beta')=0,
  \eeq
 where the norm kernel
 ${\cal N}^J_{n }(\beta,\beta')$
 and Hamiltonian kernel
 ${\cal H}^J_{n }(\beta,\beta')$
 are defined as
 \beq
 {\cal O}^J_{n}(\beta,\beta')\equiv
  \langle \Phi^{(N\Lambda)}_{n}(\beta) \vert \hat O
 \hat{P}^J_{KK}\hat{P}^N \hat{P}^Z \vert\Phi^{(N\Lambda)}_{n}
 (\beta')\rangle
 \eeq
 with $\hat{O}=1$ and $\hat{O}=\hat{H}$, respectively. The solution of the HWG equation (\ref{HWE}) provides the energy
 $E^{J}_{n\alpha}$ and weight function $f^{J}_{n\alpha}(\beta)$ for the low-lying states of hypernuclei.
 Because we begin with an energy functional rather than a Hamiltonian, we replace the  Hamiltonian overlap
 with the energy functional in which the diagonal densities and currents are replaced with mixed ones~\cite{Yao09,Yao10}.

 We note that this framework has been applied to the low-lying nuclear states if the reference
 states in Eq.~(\ref{Refst}) are from the RMF-PC calculation for nuclei~\cite{Yao10,Yao14}.

 \section{Numerical Details} \label{Sec.III}

 In the RMF-PC calculation, parity, $x$-simplex symmetry, and time-reversal invariance are imposed.
 The densities are invariant under the reflection with respect to the three planes $x$-$y$, $x$-$z$, and $y$-$z$.
 The Dirac equation Eq.~(\ref{Dirac:NL}) is solved by expanding the large and small components of the Dirac spinors $\psi^{B}_k(\br)$ separately on the basis of eigenfunctions of a three-dimensional harmonic oscillator in Cartesian coordinates with ten major shells which are found to be sufficient for the hypernuclei under consideration. The mass of the $\Lambda$ hyperon is taken as $m_\Lambda= 1115.6$ MeV$/c^2$. Pairing correlation between the nucleons is treated with the BCS approximation by using a density-independent $\delta$ force with a smooth cut off factor~\cite{Krieger90}.

 In the projection calculation, the Gauss-Legendre quadrature is used for the integral over Euler angle $\theta$. The number of mesh points in the interval $[0, \pi]$ for the Euler angle $\theta$ and gauge angle $\varphi_{\tau}$ is chosen as 14 and 9 in the angular momentum and particle number projection, respectively. The Pfaffian method~\cite{Robledo09} is applied to evaluate the phase of the norm overlap in the kernels.

 %
 \section{Results and discussions}\label{Sec.IV}
 \subsection{Mean-field calculation}\label{Sec.A}
 \subsubsection{Hyperon impurity effect}
 %

 \begin{figure}[tbp]
 \centering
 \includegraphics[width=7.6cm]{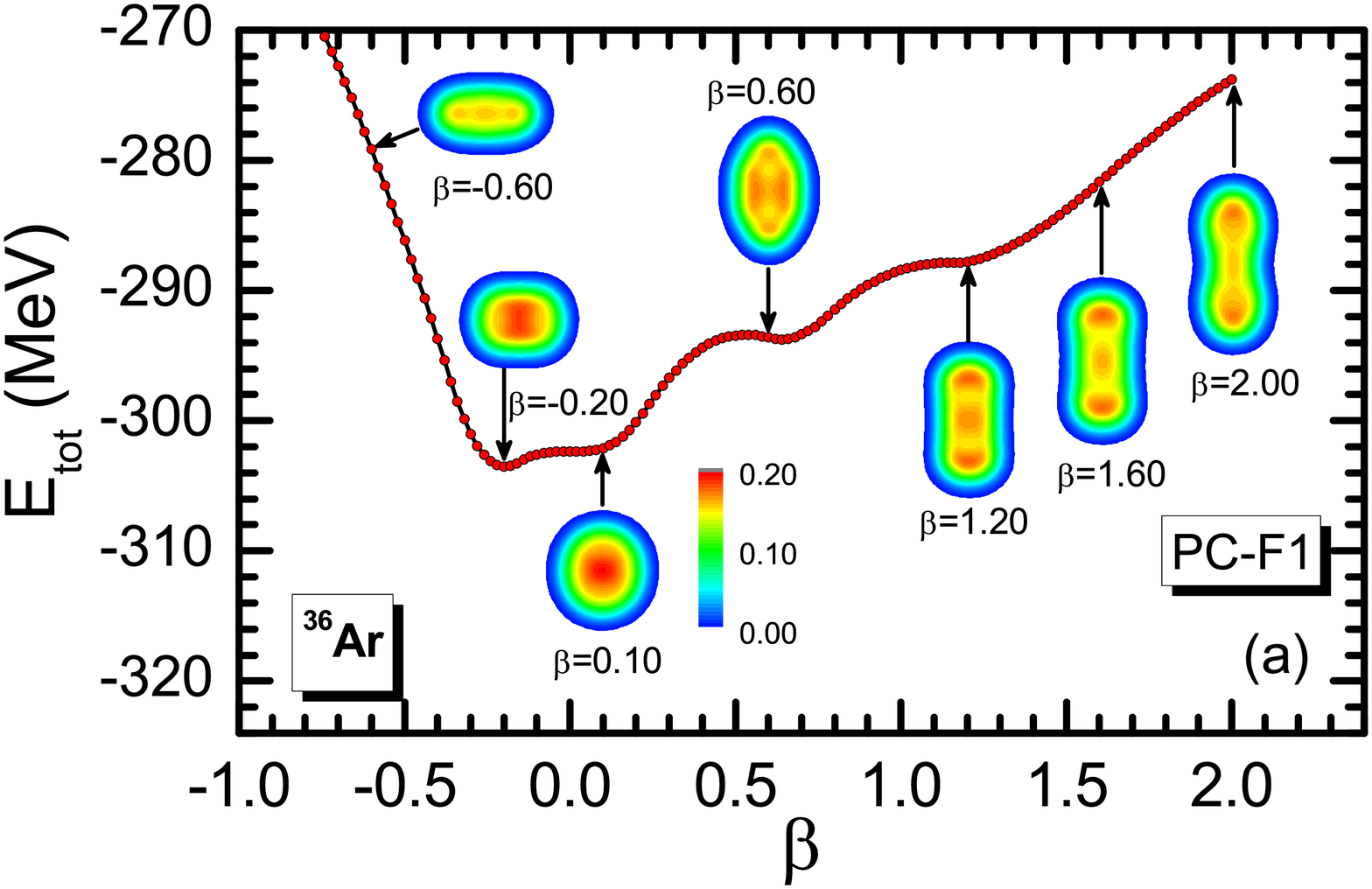}\vspace{-0.05cm}
 \includegraphics[width=7.6cm]{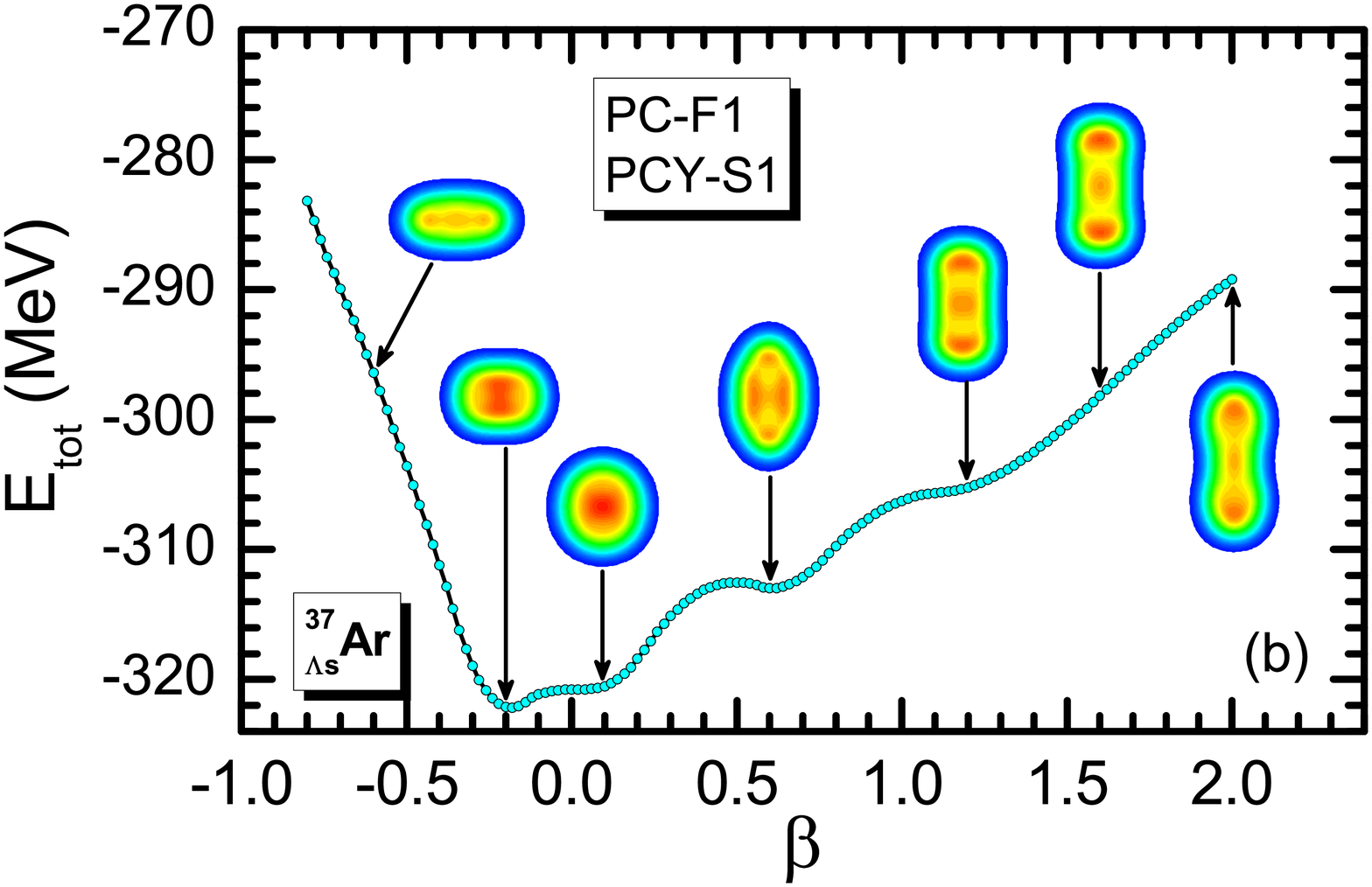}\vspace{-0.05cm}
 \includegraphics[width=7.6cm]{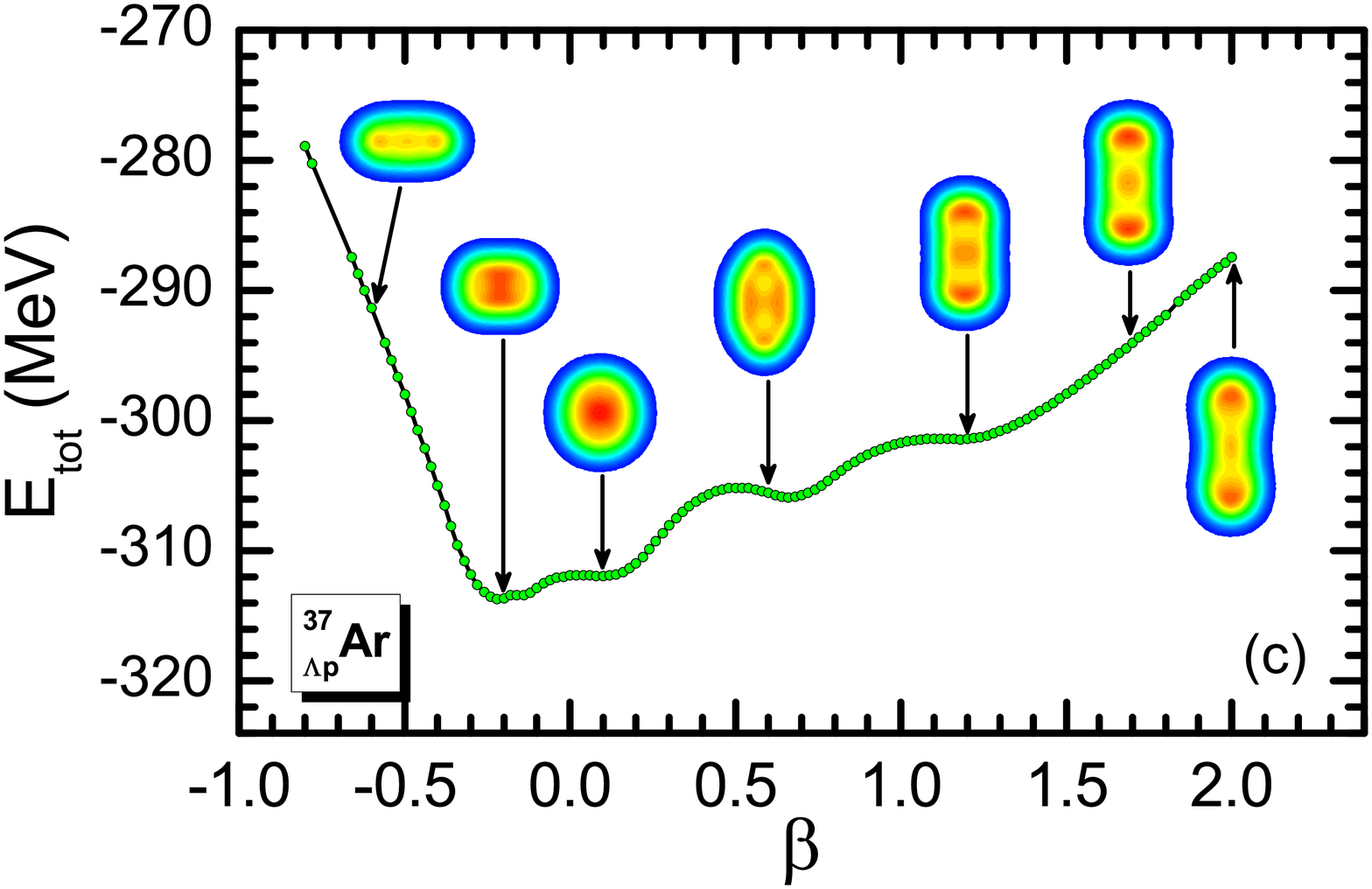}\vspace{-0.05cm}
 \includegraphics[width=7.6cm]{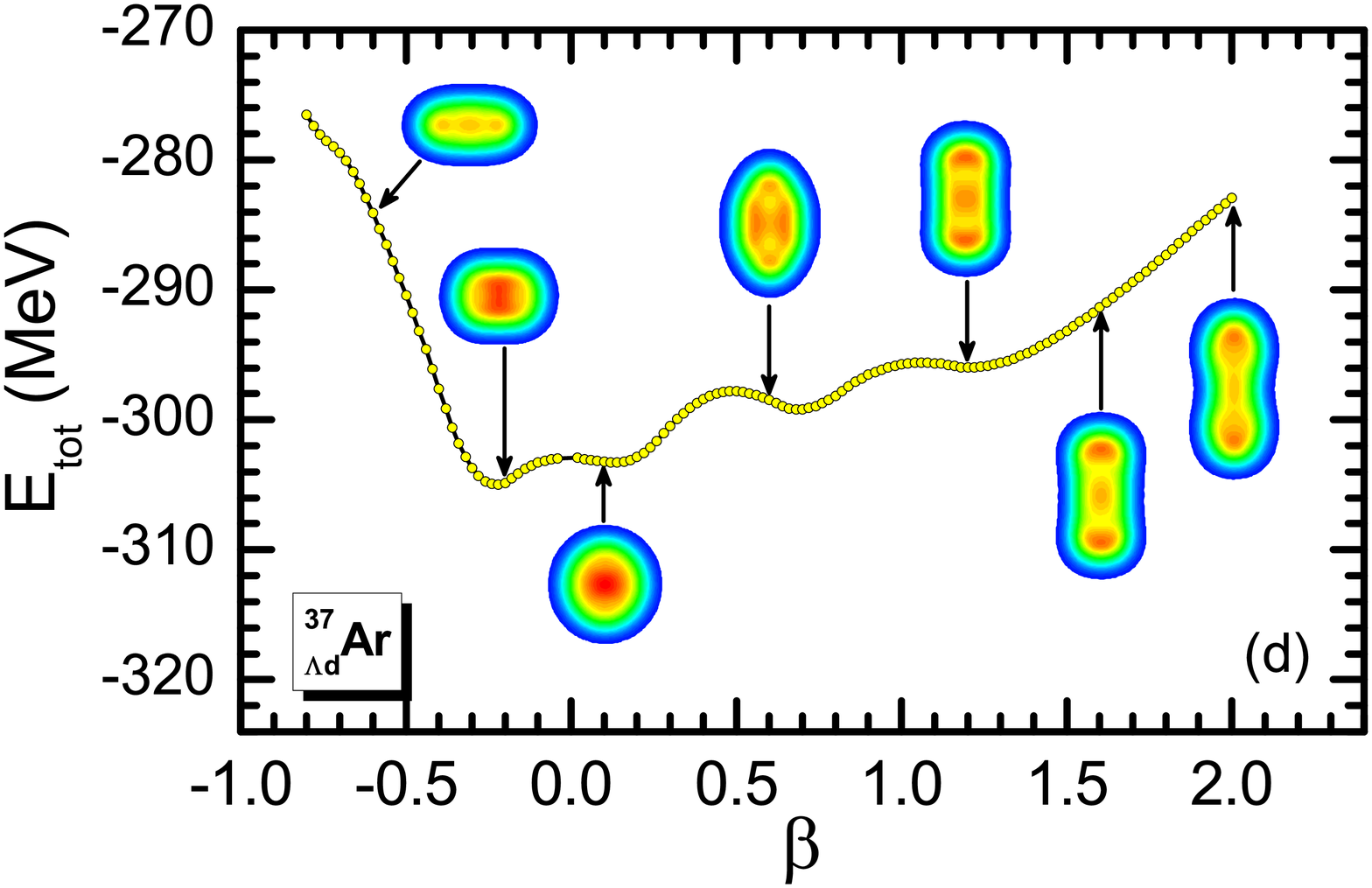}
 \caption{
  (Color online) The total energy of (a) $^{36}$Ar and (b) $^{37}_{\Lambda_s}$Ar,
  (c) $^{37}_{\Lambda_p}$Ar, and (d) $^{36}_{\Lambda_d}$Ar as a function of the quadrupole deformation parameter $\beta$. The parameter sets PC-F1 and PCY-S1 are adopted for the $NN$ and $N\Lambda$ effective interactions, respectively. The insets are the contours of the nuclear intrinsic density distributions in the $y$-$z$ plane at $x=0$ fm corresponding to some points in the curves. }
 \label{fig:PECsAr36}
 \end{figure}

 Figure~\ref{fig:PECsAr36} displays the total energies of $^{36}$Ar and $^{37}_{\Lambda_s}$Ar, $^{37}_{\Lambda_p}$Ar, and $^{37}_{\Lambda_d}$Ar as a function of the quadrupole deformation parameter $\beta$ with the PC-F1 ($NN$) + PCY-S1 ($N\Lambda$) parameter set. The hyperon is put in the lowest one of the states which correspond to the $s, p$, or $d$ state in the spherical limit, respectively. The density profiles for some selected deformed configurations are also plotted in Fig.~\ref{fig:PECsAr36}. The mean-field energy curves are similar to those in the RMF-ME model~\cite{Lu14}.

 One can see a global minimum of the binding energy of $^{36}$Ar located at the oblate shape with $\beta\simeq-0.20$ and a shallow SD minimum at $\beta\simeq0.64$ with the excitation energy $E_x=9.8$ MeV. Considering the triaxial $\gamma$ deformation, the shallow minima or shoulders around $\beta\sim0.1$ and $\beta\sim1.2$ turn out to be actually two saddle points of the energy surface in $\beta$-$\gamma$ plane. With the addition of one hyperon in $s, p,$ or $d$ state, respectively, the topography of the energy curve does not change dramatically. The global oblate minimum and the SD minimum persist in $^{37}_\Lambda$Ar. Quantitatively, the deformation of the global minimum is slightly decreased to $-0.18$ in $^{37}_{\Lambda_s}$Ar, while that of the SD minimum becomes $0.60$. In contrast, the deformation parameter of global minimum and SD minimum is increased to $\beta=-0.22$ and $\beta=0.66$, respectively, in $^{37}_{\Lambda_p}$Ar. For $^{37}_{\Lambda_d}$Ar, these values are $\beta=-0.22$ and $\beta=0.68$, respectively. The shape-driving effects of the $\Lambda_s, \Lambda_p,$ and $\Lambda_d$ in $^{37}_{\Lambda}$Ar are consistent with the findings for other $sd$-shell nuclei demonstrated in our previous investigation~\cite{Xue15}.

 \begin{table*}[t]
 \caption{
 \label{table1}
 The quadrupole deformation parameters ($\beta_{2}$, $\beta_{\Lambda}$), rms radii of hypernuclei ($R_m$), neutrons ($R_n$), protons ($R_p$), and the hyperon ($R_\Lambda$), and the proton skin ($\Delta R_{pn}\equiv R_p-R_n$) for the normal deformed (ND) and superdeformed (SD) states of $^{36}$Ar and $^{37}_{\Lambda_s}$Ar, $^{37}_{\Lambda _p}$Ar, and $^{37}_{\Lambda _d}$Ar from mean-field calculation. The excitation energies ($E_x$) for the SD are also calculated. }
 \begin{ruledtabular}
 \centering
 \begin{tabular}{cccccccccccccccccccccccccccc}
 &     &    \multicolumn{7}{c}{Normal deformed (ND) states}    & \multicolumn{8}{c}{Superdeformed (SD) states }  \\
  \cline{3-9\  \ }  \cline{10-17\  \ }
 &     &   \multicolumn{2}{c}{Deformation}    &\multicolumn{4}{c}{rms radii (fm)} &\multicolumn{1}{c}{skin (fm)} & \multicolumn{2}{c}{Deformation} & \multicolumn{4}{c}{rms radii (fm)} &\multicolumn{1}{c}{skin (fm)} &\multicolumn{1}{c}{$E$ (MeV) }  \\
  \cline{3-4\  \ } \cline{5-8\  \ }  \cline{9-9\  \ } \cline{10-11\  \ } \cline{12-15\  \ } \cline{16-16\  \ }   \cline{17-17\  \ }
 Parameter  & Nucleus & $\beta_{2}$ & $\beta_{\Lambda}$ & $~R_m$ & $R_n$ &  $R_p$  & $R_\Lambda$ & $~~\Delta R_{pn}$ & $\beta_{2}$ & $\beta_{\Lambda}$ & $~R_m$ & $R_n$ &  $R_p$ & $R_\Lambda$ & $\Delta R_{pn}$ & ~$E_x$~ \\
 \hline
                             PC-F1             &$            {}^{36}$Ar   &-0.20  & $$      & 3.278 &  3.257  &  3.299  &       & 0.042  & 0.64   &
 & 3.403    &  3.382   &  3.425   &       & 0.043  & 9.786 \\ \\

 $\multirow{3}{*}{\centering  PC-F1 } $ &$_{\Lambda_s}^{37}$Ar   &-0.18  & -0.016  & 3.256 &  3.249  &  3.291  & 2.686 & 0.042  & 0.60   & 0.165
 & 3.366    &  3.362   &  3.404   & 2.671 & 0.042  & 9.188 \\
 $\multirow{3}{*}{\centering  PCY-S1 }$ &$_{\Lambda _p}^{37}$Ar$$ &-0.22  & -0.439  & 3.287 &  3.263  &  3.305  & 3.402 & 0.042  & 0.66   & 1.232
 & 3.411    &  3.384   &  3.426   & 3.615 & 0.042  & 7.814 \\
                                               &$_{\Lambda _d}^{37}$Ar$$ &-0.22  & -0.879  & 3.307 &  3.259  &  3.301  & 4.178 & 0.042  & 0.68   & 1.983
 & 3.428    &  3.384   &  3.425   & 4.172 & 0.041  & 5.802 \\ \\
 $\multirow{3}{*}{\centering  PC-F1 } $ &$_{\Lambda_s}^{37}$Ar   &-0.18  & -0.024  & 3.228 &  3.227  &  3.269  & 2.381 & 0.042  & 0.60   & 0.138
 & 3.343    &  3.343   &  3.386   & 2.420 & 0.043  & 9.834\\
 $\multirow{3}{*}{\centering  PCY-S2 }$ &$_{\Lambda _p}^{37}$Ar$$ &-0.20  & -0.381  & 3.258 &  3.240  &  3.282  & 3.157 & 0.042  & 0.64   & 1.037
 & 3.380    &  3.360   &  3.402   & 3.331 & 0.042  & 7.643 \\
                                               &$_{\Lambda _d}^{37}$Ar$$ &-0.22  & -0.848  & 3.299 &  3.250  &  3.292  & 4.172 & 0.042  & 0.68   & 1.834
 & 3.413    &  3.375   &  3.416   & 3.987 & 0.041  & 5.947 \\ \\
 $\multirow{3}{*}{\centering  PC-F1 } $ &$_{\Lambda_s}^{37}$Ar   &-0.18  & -0.014  & 3.256 &  3.249  &  3.291  & 2.686 & 0.042  & 0.60   & 0.163
 & 3.365    &  3.362   &  3.403   & 2.670 & 0.041  & 9.168 \\
 $\multirow{3}{*}{\centering  PCY-S3 }$ &$_{\Lambda _p}^{37}$Ar$$ &-0.18  & 0.166   & 3.267 &  3.247  &  3.289  & 3.234 & 0.042  & 0.66   & 1.211
 & 3.410    &  3.384   &  3.425   & 3.614 & 0.041  & 7.889 \\
                                               &$_{\Lambda _d}^{37}$Ar$$ &-0.22  &-0.862   & 3.303 &  3.257  &  3.299  & 4.122 & 0.042  & 0.68   & 1.944
 & 3.427    &  3.384   &  3.425   & 4.171 & 0.041  & 6.368 \\ \\
 $\multirow{3}{*}{\centering  PC-F1 } $ &$_{\Lambda_s}^{37}$Ar   &-0.18  & -0.045  & 3.246 &  3.243  &  3.285  & 2.490 & 0.042  & 0.60   & 0.173
 & 3.357    &  3.357   &  3.399   & 2.497 & 0.042  & 9.497 \\
 $\multirow{3}{*}{\centering  PCY-S4 }$ &$_{\Lambda _p}^{37}$Ar$$ &-0.20  & -0.406  & 3.272 &  3.251  &  3.293  & 3.268 & 0.042  & 0.66   & 1.097
 & 3.404    &  3.383   &  3.424   & 3.420 & 0.041  & 7.554 \\
                                               &$_{\Lambda _d}^{37}$Ar$$ &-0.22  & -0.874  & 3.307 &  3.259  &  3.301  & 4.167 & 0.042  & 0.68   & 1.951
 & 3.426    &  3.383   &  3.424   & 4.161 & 0.041  & 6.162 \\
 \end{tabular}
 \end{ruledtabular}
 \end{table*}

 \begin{figure}[tbp]
 \centering
 \includegraphics[clip=,width=8.5cm]{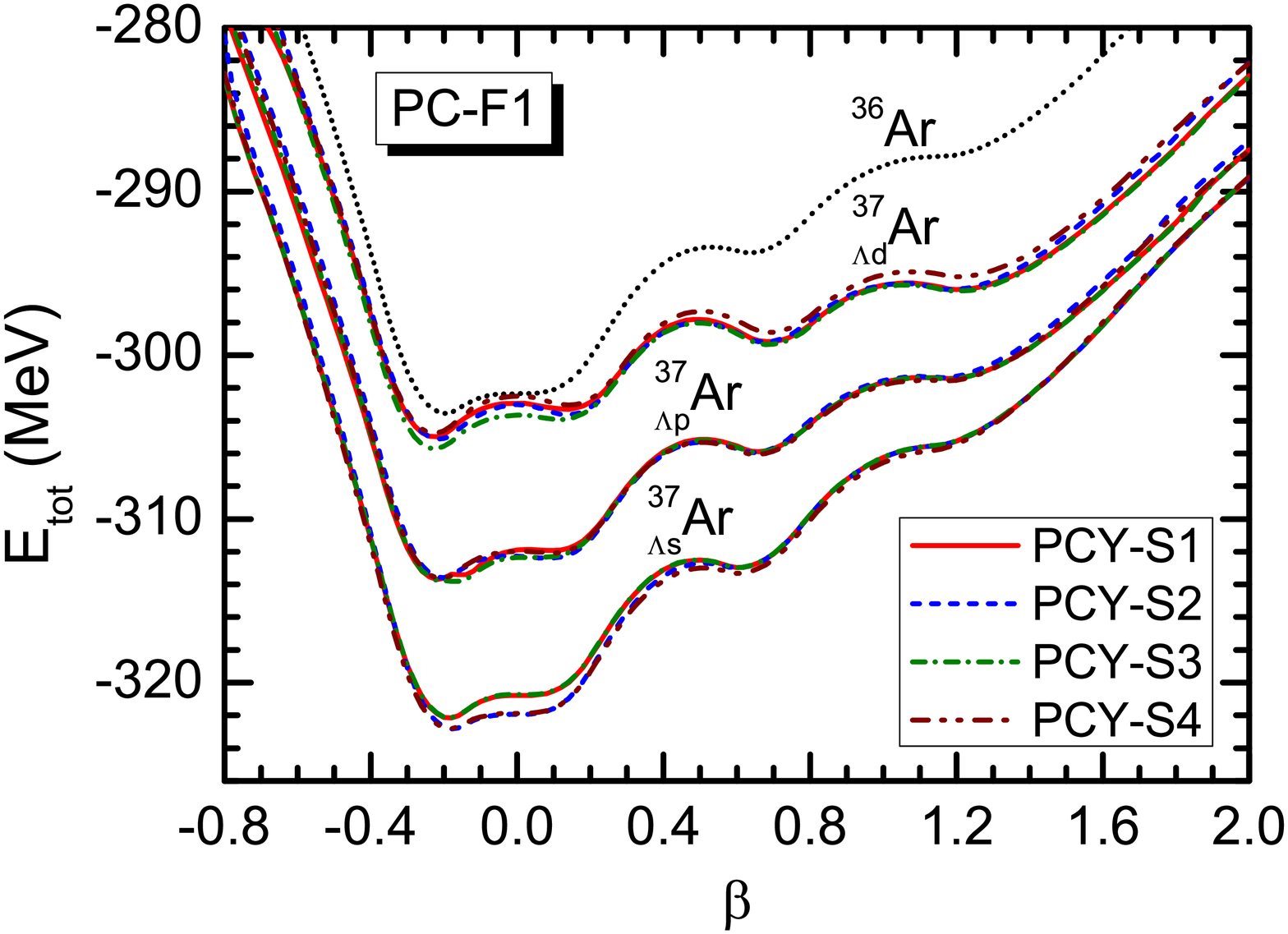}
 \caption{(Color online) The potential energy curves (PECs) of $^{37}_{\Lambda_s}$Ar, $^{37}_{\Lambda_p}$Ar, and $^{36}_{\Lambda_d}$Ar, calculated by the four $N\Lambda$ interactions, respectively, as a function of the quadrupole deformation parameter $\beta$. The PEC of $^{36}$Ar is also shown for comparison.}
 \label{fig:PECsAr37}
 \end{figure}

 To investigate the force-parameter dependence of the results, we perform the calculation with the other three effective $N\Lambda$ interactions PCY-S2, PCY-S3, and PCY-S4, respectively. The potential energy curves (PECs) of $^{37}_{\Lambda_s}$Ar, $^{37}_{\Lambda_p}$Ar, and $^{36}_{\Lambda_d}$Ar are exhibited in Fig.~\ref{fig:PECsAr37}. Similar topographies of the PECs are shown for the four $N\Lambda$ sets, respectively, when the $\Lambda$ is put in the same orbital. The SD states persist in $^{37}_{\Lambda}$Ar for all four effective interactions. A small difference amongst the predictions of the four interactions is shown in the region around spherical shape. The detailed information of the predicted ND and SD states in $^{37}_{\Lambda}$Ar is listed in Table~\ref{table1}. All the four interactions predict rather similar deformations for the ND and SD configurations. Besides, we note that for all the $N\Lambda$ interactions except for the PCY-S2 interaction, the predicted excitation energy of the SD state in $^{37}_\Lambda$Ar is lower than that in $^{36}$Ar. In particular, the excitation energy decreases from 9.2 MeV to 5.8 MeV in the hypernucleus from $^{37}_{\Lambda s}$Ar to $^{37}_{\Lambda d}$Ar for the PCY-S1 . The shrinkage effect of $\Lambda_s$ on nuclear size is also shown in Table~\ref{table1}. However, the $\Lambda_p$ and $\Lambda_d$ may either increase or decrease the rms radii of neurons and protons, depending on the details of the effective $N\Lambda$ interaction. In particular, the PCY-S2 predicts the rms radii of neutrons ($R_n$), protons ($R_p$) and the hyperon ($R_\Lambda$) are much smaller than the other three $N\Lambda$ interactions. One may understand it as a consequence of the missing $N\Lambda$ tensor coupling term in the PCY-S2. Of particular interest is that the proton skin $\Delta R_{pn}\equiv R_p-R_n$ is not changed at all by one $\Lambda$, as a consequence of the isoscalar nature of a $\Lambda$ hyperon.

 \begin{table*}[t]
 \caption{
 \label{table2}
 The quadrupole deformation parameters ($\beta_{2}$, $\beta_{\Lambda}$), rms radii of baryons ($R_m$) and the $\Lambda$ ($R_\Lambda$), total energies ($E_\ttot$, $E_{\rm exp}$), single-$\Lambda$ separation energy ($S_{\Lambda}$), and the overlap ($I_\tovlp$) between $\Lambda$ hyperon and the nucleons in the core for the normal deformed (ND) and superdeformed (SD) [labeled by asterisks] states of $^{36}$Ar and $^{37}_{\Lambda_s}$Ar, in comparison with the results from the other models.}
 \begin{ruledtabular}
 \begin{tabular}{ccccccccccccccccccc}
 &     &    \multicolumn{2}{c}{Deformation}   &\multicolumn{2}{c}{rms radii (fm)}   & \multicolumn{3}{c}{Energies (MeV)}  & \multicolumn{1}{c}{Overlap (fm$^{-3}$) } \\
  \cline{3-4\  \ } \cline{5-6\  \ } \cline{7-9\  \ }\cline{10-10\  \ }
 Model & Nucleus &  $\beta_{2}$ & $\beta_{\Lambda}$ &  $~~R_m$    & $R_\Lambda$ & $~E_{\rm tot}$ & $E_{\rm exp}$   & $S_{\Lambda}$  &~$I_\tovlp$~~ \\
 \hline
 $\multirow{3}{*}{\centering RMF-PC}$           &$        {}^{36}$Ar        &-0.200   & $$        & 3.278     &           & -303.540  &-306.716    &         &          \\
 $\multirow{3}{*}{\centering (PC-F1, PCY-S1)}$   &$        {}^{36}$Ar$^\ast$ &0.640    & $$        & 3.403     &           & -293.754  &$$          &         &          \\
                                                &$_{\Lambda_s}^{37}$Ar      &-0.180   & -0.016    & 3.256     &  2.686    & -322.154  &$$          & 18.614  & 0.1323   \\
                                                &$_{\Lambda_s}^{37}$Ar$^\ast$ &0.600    & 0.165     & 3.366     &  2.671    & -312.966  &$$          & 19.212  & 0.1338   \\
 \hline
 $\multirow{3}{*}{\centering RMF-PC}$           &$        {}^{36}$Ar        &-0.180   & $$        & 3.252     &           & -303.659  &-306.716    &         &         \\
 $\multirow{3}{*}{\centering (PC-PK1, PCY-S1)}$  &$        {}^{36}$Ar$^\ast$ &0.600    & $$        & 3.352     &           & -295.731  &$$          &         &         \\
                                                &$_{\Lambda_s}^{37}$Ar      &-0.160   & 0.014     & 3.234     &  2.725    & -321.733  &$$          & 18.074  & 0.1337 \\
                                                &$_{\Lambda_s}^{37}$Ar$^\ast$ &0.560    & 0.144     & 3.319     &  2.694    & -314.575  &$$          & 18.844  & 0.1368  \\
 \hline
 $\multirow{3}{*}{\centering RMF-ME~\cite{Lu14}}$         &$        {}^{36}$Ar        &-0.212   & $$        & 3.238     &            & -303.802  &-306.716    &         &         \\
 $\multirow{3}{*}{\centering (PK1, PK1-Y1)}$    &$        {}^{36}$Ar$^\ast$ &0.620    & $$        & 3.346     &            & -296.670  &$$          &         &          \\
                                               &$_{\Lambda_s}^{37}$Ar      &-0.204   & -0.057    & 3.220     &  2.644     & -321.979  &$$          & 18.177  & 0.1352  \\
                                               &$_{\Lambda_s}^{37}$Ar$^\ast$ &0.597    & 0.172     & 3.319     &  2.626     & -315.194  &$$          & 18.524  & 0.1370  \\
 \hline
 $\multirow{3}{*}{\centering HyperAMD~\cite{Isaka15}}$             &$        {}^{36}$Ar          &-0.21    & $$        &           &           & -301.06    &-306.716  &         &         \\
 $\multirow{3}{*}{\centering (D1S, YNG-ESC08c)}$&$        {}^{36}$Ar$^\ast$   & 0.65    & $$        &           &           & -291.77    &$$        &         &         \\
                                               &$_{\Lambda_s}^{37}$Ar        &-0.19    & -0.07     &           &           & -319.64    &$$        & 18.59   & 0.1338  \\
                                               &$_{\Lambda_s}^{37}$Ar$^\ast$   &0.64     & 0.20      &           &           & -309.81    &$$        & 18.04   & 0.1310  \\
 \hline
 $\multirow{3}{*}{\centering SHF~\cite{Zhou16}}$             &$        {}^{36}$Ar          &-0.170   & $$        & 3.282     &           & -304.091   &-306.716  &         &         \\
 $\multirow{3}{*}{\centering (SkI4, NSC89)}$   &$        {}^{36}$Ar$^\ast$   &0.517    & $$        & 3.417     &           & -296.418   &$$        &         &         \\
                                               &$_{\Lambda_s}^{37}$Ar        &-0.165   & -0.106    & 3.261     &  2.719    & -321.384   &$$         & 17.293  & 0.1299  \\
                                               &$_{\Lambda_s}^{37}$Ar$^\ast$   &0.515    & 0.323     & 3.397     &  2.781    & -313.540    &$$        & 17.122  & 0.1284  \\
 \end{tabular}
 \end{ruledtabular}
 \end{table*}

 \subsubsection{Correlation between $\Lambda$ separation energy and density overlap}

 The contribution to $\Lambda$ separation (or binding) energy can be divided into kinetic energy and interaction energy between the $\Lambda$ and the core nuclei.
 The contribution from the kinetic energy to the difference in the $\Lambda$ binding energies of the ND and SD states, defined as $\Delta E^{NS}_\Lambda=S^{\rm ND}_\Lambda-S^{\rm SD}_\Lambda$,
 can be roughly neglected. Therefore, several authors~\cite{Lu14,Isaka15,Zhou16} tried to understand the $\Delta E^{NS}_\Lambda$ from the interaction energy which is
 approximately proportional to the overlap $I_\tovlp$ between the densities of the core nuclei and ${\Lambda}$ hyperon, c.f. Eq.~(\ref{Energy:NL}),
 \beq
   \label{ovlp}
   I_\tovlp =\int d^3r \rho^\Lambda(\br)\rho^N(\br).
 \eeq

 This quantity has previously been adopted to study the triaxial deformation $\gamma$ effect on the $\Lambda$ binding energy with SHF+BCS method in Ref.~\cite{Win11}.

 \begin{figure}[tbp]
 \centering
 \includegraphics[clip=,width=8.5cm]{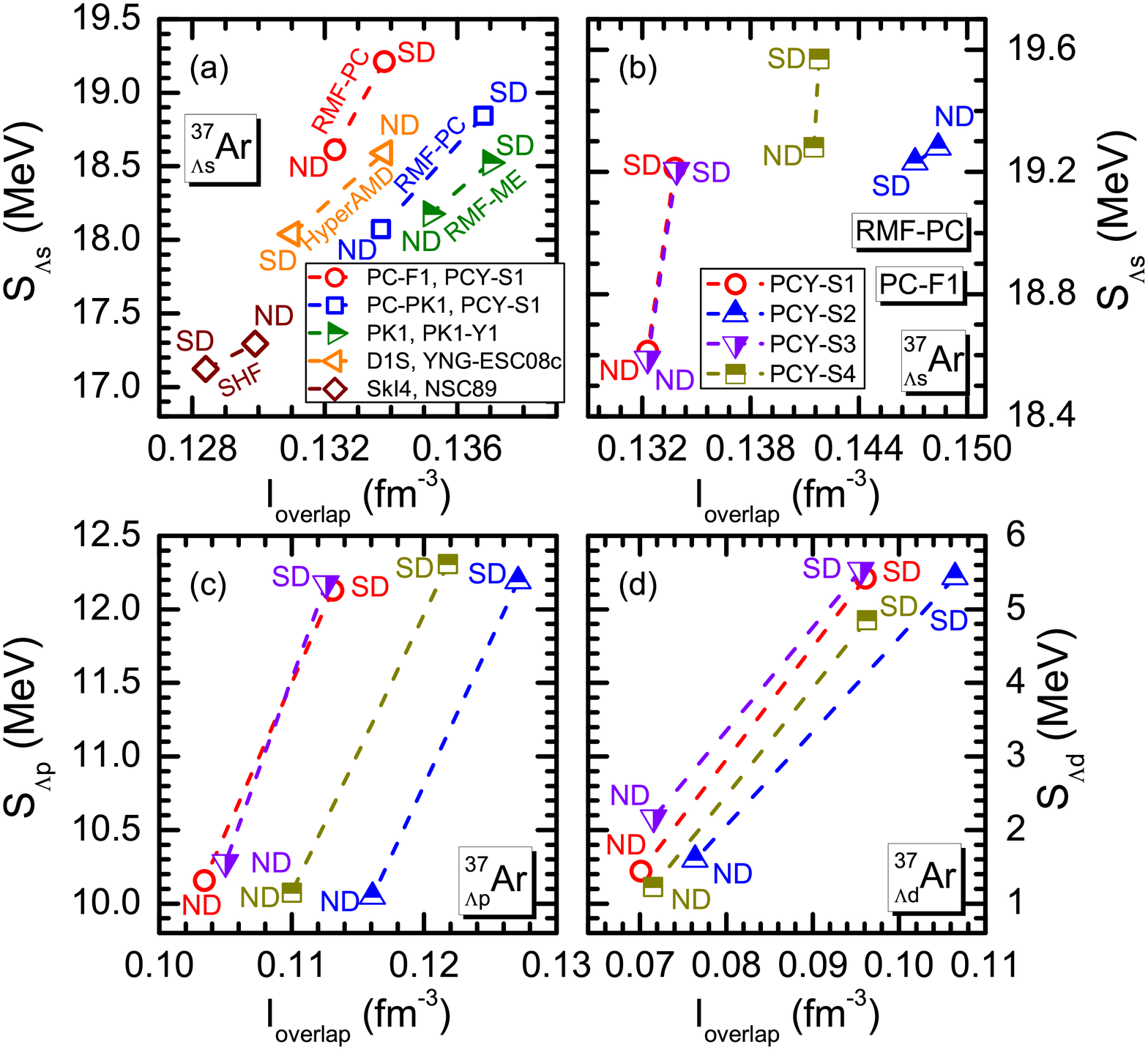}\vspace{-0cm}
 \caption{(Color online) (a) The correlation between $\Lambda$ separation energy $S_\Lambda$ and density overlap $I_\tovlp$ obtained with different
          models for the normal deformed (ND) and superdeformed (SD) states of $^{37}_{\Lambda s}$Ar, respectively. (b), (c), and (d) The correlations calculated by RMF-PC model with the $NN$
          interaction PC-F1 and four $N\Lambda$ interactions for the ND and SD states of $^{37}_{\Lambda s}$Ar, $^{37}_{\Lambda p}$Ar, and $^{37}_{\Lambda d}$Ar, respectively. }
 \label{fig:CORRE}
 \end{figure}

 Figure~\ref{fig:CORRE}(a) displays the correlation between the $\Lambda$ separation energy $S_\Lambda$ and the density overlap $I_\tovlp$ with different interactions in different models for the ND and SD states of $^{37}_{\Lambda s}$Ar. One can see that a larger $I_\tovlp$ value corresponds to a larger $S_\Lambda$. Our results show that the $S_\Lambda$ in SD states is larger than that in ND states, which is consistent with the prediction by the RMF-ME~\cite{Lu14}, but contradicts to the results from the HyperAMD~\cite{Isaka15} and SHF~\cite{Zhou16} models. The correlations between $S_\Lambda$ and $I_\tovlp$, calculated by the four $N\Lambda$ interactions for the ND and SD states of $^{37}_{\Lambda s}$Ar, $^{37}_{\Lambda p}$Ar, and $^{37}_{\Lambda d}$Ar, are shown in the Fig.~\ref{fig:CORRE}(b), (c), and (d), respectively. One observes that the $S_\Lambda$ and $I_\tovlp$ in SD states are always larger than these in ND states in all cases except for the PCY-S2 interaction in $^{37}_{\Lambda s}$Ar.

 Table~\ref{table2} lists the deformation parameters, rms radii, total energy, $\Lambda$ separation energy, and  the $I_\tovlp$ for both the ND and SD states in $^{36}$Ar and $^{37}_{\Lambda s}$Ar, in comparison with the results of other models.
 The results of all the models are rather similar. However, if one analyzed the results in a quantitative way, one can see the following points.
 \begin{itemize}
 \item The change of the deformation for both the ND and SD states induced by the $\Lambda$ in the relativistic models is significantly larger than that in the non-relativistic models. This point has already been discussed in Refs.~\cite{Schulze10,Xue15}.
 \item The SHF model predicted the smallest deformation, $\Lambda$ separation energy and $I_\tovlp$ for both the ND and SD hypernuclear states. For the latter two, it may have something to do with the fact that the mean-field potentials in the SHF model are shallower than those of the RMF models~\cite{Zhou16}.
 \item The $\Lambda_s$ separation energy in the SD state is predicted to be larger than that in the ND state in the relativistic models. However, an controversial results was pointed out in non-relativistic models. We note that the overlap $I_\tovlp$ between the $\Lambda$ hyperon and core nuclei is correlated to the $\Lambda$ separation energy. This correlation is further investigated with different sets of $N\Lambda$ interaction for $^{37}_{\Lambda s}$Ar, $^{37}_{\Lambda p}$Ar and $^{37}_{\Lambda d}$Ar, respectively, as illustrated in Table~\ref{table3}. In particular, one finds from Table~\ref{table3} that the $\Lambda$ separation energy of the SD state becomes increasingly larger than that of the ND state as the valence $\Lambda$ is put from  $s$ orbit to $d$ orbit. Similar conclusions are drawn for $^{49}_{\Lambda}$Ar and $^{33}_{\Lambda}$S, except that the $\Lambda_s$ separation energy of the SD states is significantly lower than that in the ND states, as demonstrated in Table~\ref{Ar-S}.
 \end{itemize}

 \begin{table*}[t]
 \caption{
 \label{table3}
 The $\Lambda$ separation energy $S_{\Lambda}$ (MeV) and overlap $I_\tovlp$ (fm$^{-3}$) calculated by four different parameter sets of $N\Lambda$ interactions for both the normal deformed (ND) and superdeformed (SD) [labeled by asterisks] states in $^{37}_{\Lambda_s}$Ar, $^{37}_{\Lambda _p}$Ar, and $^{37}_{\Lambda _d}$Ar, respectively.  }
 \begin{ruledtabular}
 \begin{tabular}{ccccccccccccc}
 &  &  \multicolumn{2}{c}{PC-F1, PCY-S1}   &\multicolumn{2}{c}{PC-F1, PCY-S2}   & \multicolumn{2}{c}{PC-F1, PCY-S3}  & \multicolumn{2}{c}{PC-F1, PCY-S4 } \\
 \cline{3-4\  \ } \cline{5-6\  \ } \cline{7-8\  \ }\cline{9-10\  \ }
 & Nucleus &  $S_{\Lambda}$  &~$I_\tovlp$ &  $S_{\Lambda}$  &~$I_\tovlp$ & $S_{\Lambda}$  &~$I_\tovlp$  & $S_{\Lambda}$  &~$I_\tovlp$~~ \\
 \hline
  &$_{\Lambda _s}^{37}$Ar        & 18.614  & 0.1323   & 19.279  & 0.1484    & 18.588  & 0.1323   & 19.281  & 0.1415   \\
  &$_{\Lambda _s}^{37}$Ar$^\ast$ & 19.212  & 0.1338   & 19.231  & 0.1471    & 19.206  & 0.1339   & 19.570  & 0.1418   \\
  &$_{\Lambda _p}^{37}$Ar        & 10.157  & 0.1034   & 10.048  & 0.1161    & 10.279  & 0.1050   & 10.075  & 0.1100   \\
  &$_{\Lambda _p}^{37}$Ar$^\ast$ & 12.129  & 0.1131   & 12.191  & 0.1271    & 12.176  & 0.1126   & 12.307  & 0.1218   \\
  &$_{\Lambda _d}^{37}$Ar        & 1.440   & 0.0701   & 1.598   & 0.0764    & 2.170   & 0.0716   & 1.228   & 0.0715  \\
  &$_{\Lambda _d}^{37}$Ar$^\ast$ & 5.424   & 0.0961   & 5.437   & 0.1065    & 5.535   & 0.0956   & 4.852   & 0.0963  \\
 \end{tabular}
 \end{ruledtabular}
 \end{table*}

 \begin{table}[t]
 \caption{The $\Lambda$ separation energies $S_{\Lambda}$ (MeV) and overlap $I_\tovlp$ (fm$^{-3}$)
 of normal deformed (ND) and superdeformed (SD) [labeled with asterisk] states for $^{49}_\Lambda$Ar and $^{33}_\Lambda$S, respectively.  }
 \begin{ruledtabular}
 \begin{tabular}{ccccccccccccccccccc}
 &  &  \multicolumn{2}{c}{PCY-S1}   &\multicolumn{2}{c}{PCY-S2}   & \multicolumn{2}{c}{PCY-S4 } \\
 \cline{3-4\  \ } \cline{5-6\  \ } \cline{7-8\  \ }
 & Nucleus &  $S_{\Lambda}$  &~$I_\tovlp$ &  $S_{\Lambda}$  &~$I_\tovlp$ & $S_{\Lambda}$  &~$I_\tovlp$  \\
 \hline
 &$_{\Lambda _s}^{49}$Ar        & 20.277  & 0.1350   & 20.640  & 0.1486    & 20.951  & 0.1426   \\
 &$_{\Lambda _s}^{49}$Ar$^\ast$ & 19.919  & 0.1323   & 19.965  & 0.1440    & 20.480  & 0.1395   \\

 &$_{\Lambda _p}^{49}$Ar        & 13.050  & 0.1148   & 12.673  & 0.1273    & 12.854  & 0.1199   \\
 &$_{\Lambda _p}^{49}$Ar$^\ast$ & 14.426  & 0.1201   & 14.437  & 0.1329    & 14.697  & 0.1278   \\

 &$_{\Lambda _d}^{49}$Ar        & 4.592   & 0.0896   & 4.524   & 0.0988    & 4.381   & 0.0901   \\
 &$_{\Lambda _d}^{49}$Ar$^\ast$ & 7.735   & 0.1021   & 7.841   & 0.1142    & 7.657   & 0.1067   \\
 \hline
 &$_{\Lambda _s}^{33}$S        & 18.570  & 0.1394   & 20.111  & 0.1617    & 19.748  & 0.1528   \\
 &$_{\Lambda _s}^{33}$S$^\ast$ & 16.625  & 0.1200   & 16.539  & 0.1337    & 17.053  & 0.1288   \\

 &$_{\Lambda _p}^{33}$S        & 9.579   & 0.1105   & 9.639   & 0.1273    & 8.797   & 0.1095   \\
 &$_{\Lambda _p}^{33}$S$^\ast$ & 11.693  & 0.1092   & 11.716  & 0.1233    & 11.959  & 0.1191   \\

 &$_{\Lambda _d}^{33}$S        & 0.309    & 0.0693   & 0.620    & 0.0776    & 0.035  & 0.0651   \\
 &$_{\Lambda _d}^{33}$S$^\ast$ & 5.363    & 0.0939   & 5.365    & 0.1052    & 4.939  & 0.0954   \\
 \end{tabular}
 \end{ruledtabular}
 \label{Ar-S}
 \end{table}

 \begin{figure}[tbp]
 \centering
 \includegraphics[clip=,width=8.5cm]{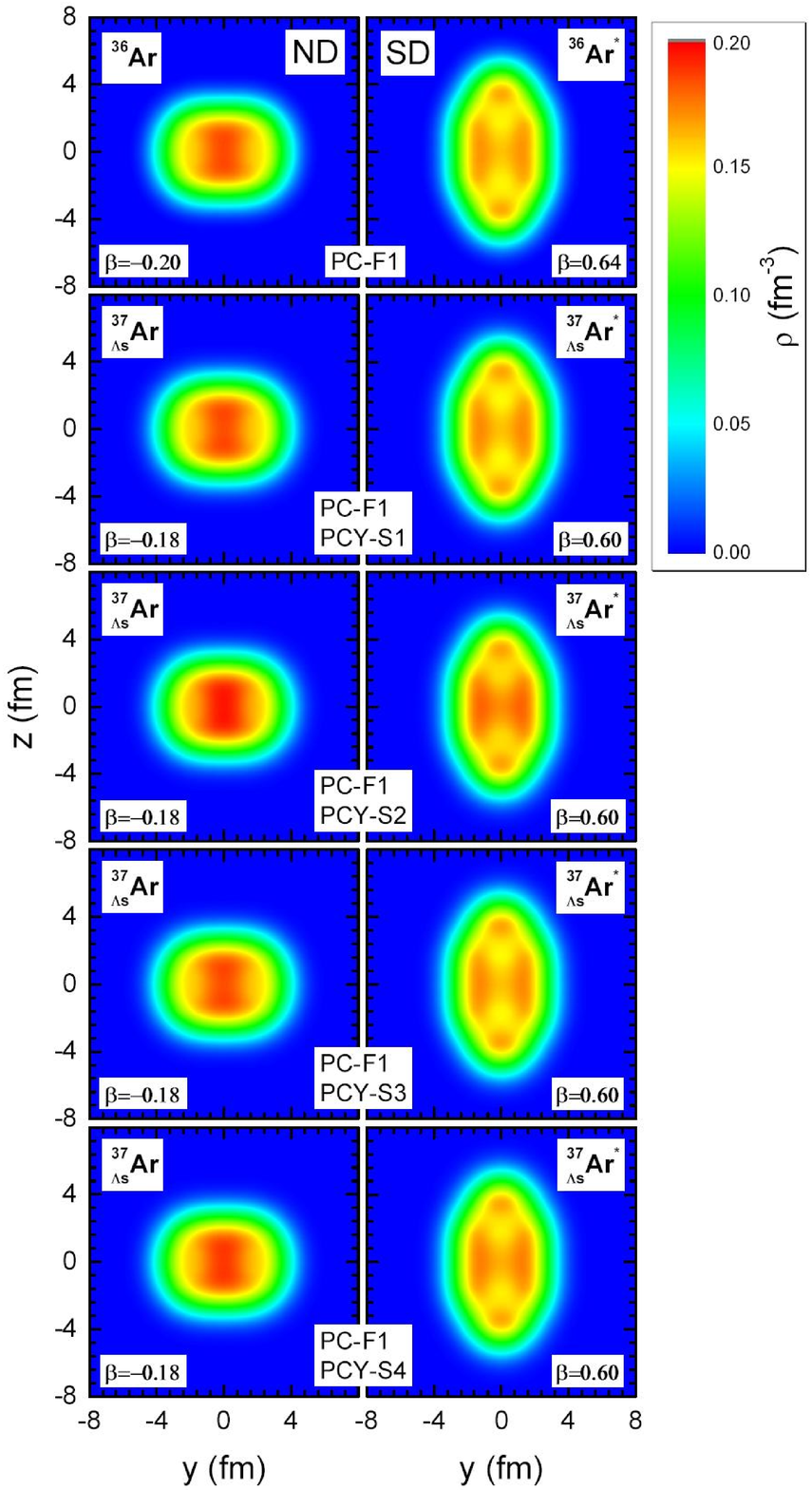}
 \caption{
  (Color online) The density distribution (in fm$^{-3}$) of total baryons in the $y$-$z$ plane
  at $x = 0$ fm (the symmetry axis is the $z$-axis) for the normal deformed (ND) and superdeformed (SD) [labeled by asterisks] states of $^{36}$Ar and $^{37}_{\Lambda _s}$Ar
  (with the four sets of $N\Lambda$ interaction, respectively).  }
 \label{fig:Ardens}
 \end{figure}

\begin{figure}[tbp]
 \centering
 \includegraphics[clip=,width=8.5cm]{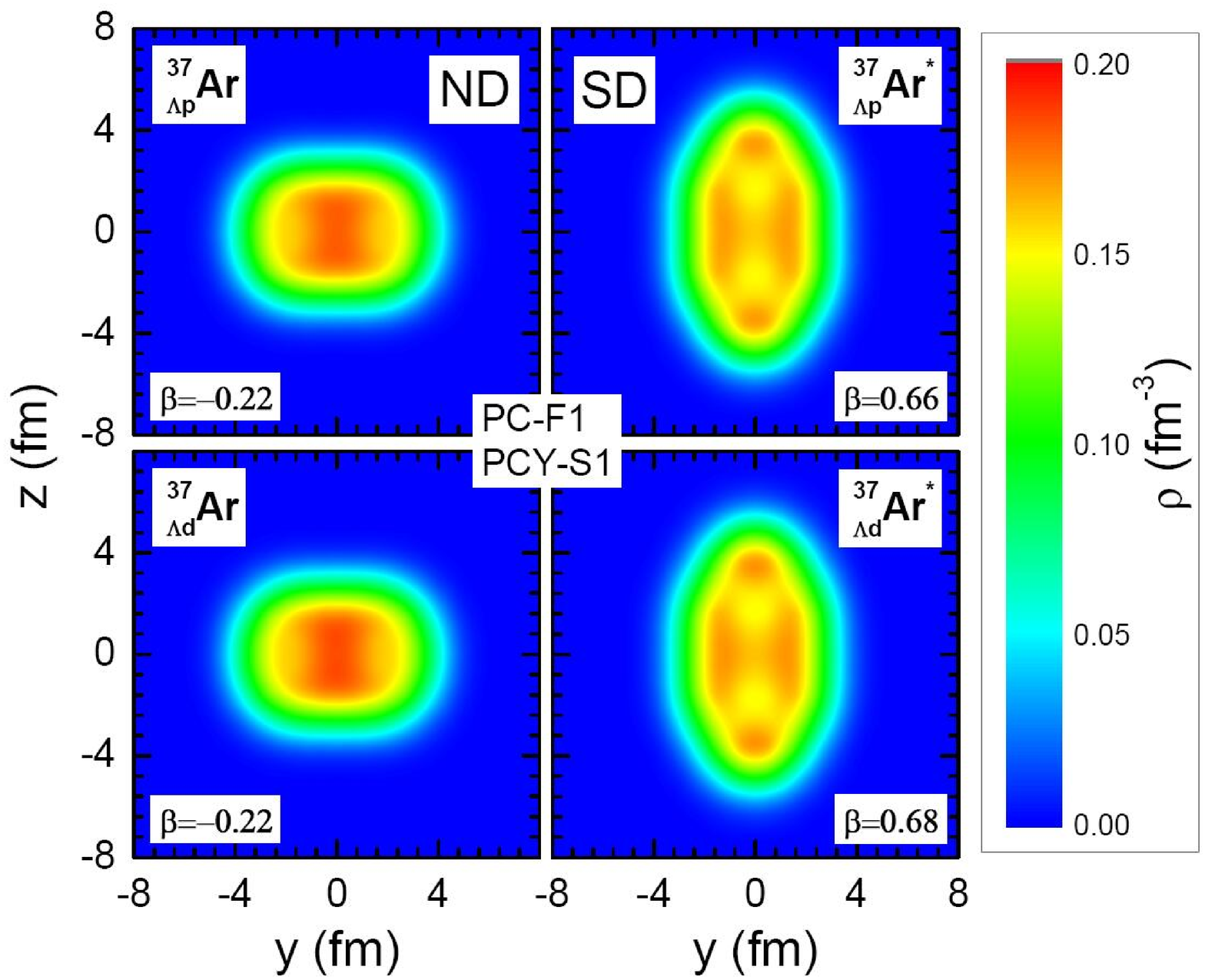}
 \caption{
  (Color online) The density distribution (in fm$^{-3}$) of total baryons in $^{37}_{\Lambda _p}$Ar and $^{37}_{\Lambda _d}$Ar in
  the $y$-$z$ plane at $x = 0$ fm (the symmetry axis is the $z$-axis). The quadrupole deformations of normal deformed (ND) and superdeformed (SD) [labeled by asterisks] states
   minima which are obtained by $NN$ interaction PC-F1 and $N\Lambda$ interaction PCY-S1 are also given.
 }
 \label{fig:Ardens-dp}
 \end{figure}

\begin{figure}[tbp]
 \centering
 \includegraphics[clip=,width=8.5cm]{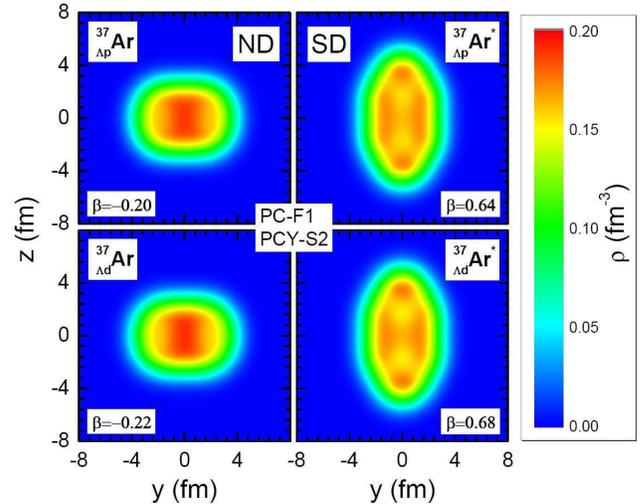}
 \caption{
  (Color online) The same as Fig.~\ref{fig:Ardens-dp}, but for the calculation with $N\Lambda$ PCY-S2 interaction.
 }
 \label{fig:Ardens-dp2}
 \end{figure}

 To shed some light on the relation between the localization of nuclear density and $\Lambda$ separation energy as suggested in Ref.~\cite{Lu14}, we plot the density distributions of baryons for both the ND and SD states of $^{36}$Ar and $^{37}_{\Lambda _s}$Ar hypernuclei with the four $N\Lambda$ interactions, respectively, as shown in Fig.~\ref{fig:Ardens}. One can see that the ring-shaped clustering structure in the SD state for all the $N\Lambda$ interactions is much less pronounced than that predicted in the RMF-ME model~\cite{Lu14}. Instead, the result is closer to that found in the HyperAMD~\cite{Isaka15} and SHF~\cite{Zhou16} models. The distributions of baryons in $^{37}_{\Lambda _p}$Ar and $^{37}_{\Lambda _d}$Ar for the PCY-S1 and PCY-S2 interactions are plotted in Fig.~\ref{fig:Ardens-dp} and Fig.~\ref{fig:Ardens-dp2}, respectively. Again, the ring-shape nuclear clustering structure is not clearly exhibited. This finding indicates that the predicted larger $\Lambda$ separation energy of SD state is not necessary attributed to the ring-shaped clustering structure of nucleons in hypernuclei. Instead, the distribution of the hyperon which depends on the details of the $N\Lambda$ interaction may play a more important role, as indicated by the behavior of the $\Delta E^{NS}_\Lambda$ for different orbital $\Lambda$, c.f. Table~\ref{table3} and Table~\ref{Ar-S}.

 In short, we find that the $\Lambda_p$ and $\Lambda_d$ binding energies of SD state are always larger than those of the ND state. However, for the $\Lambda_s$, the conclusion depends on the details of the effective nucleon-hyperon interaction and the core nuclei. We note that these conclusions are drawn based on the mean-field model. The beyond-mean-field effect may play an important role. It will be discussed in the next subsection.

 \subsection{Beyond-mean-field effect}
 \label{Sec.B}

 \begin{figure}[tbp]
 \centering
 \includegraphics[clip=,width=8cm]{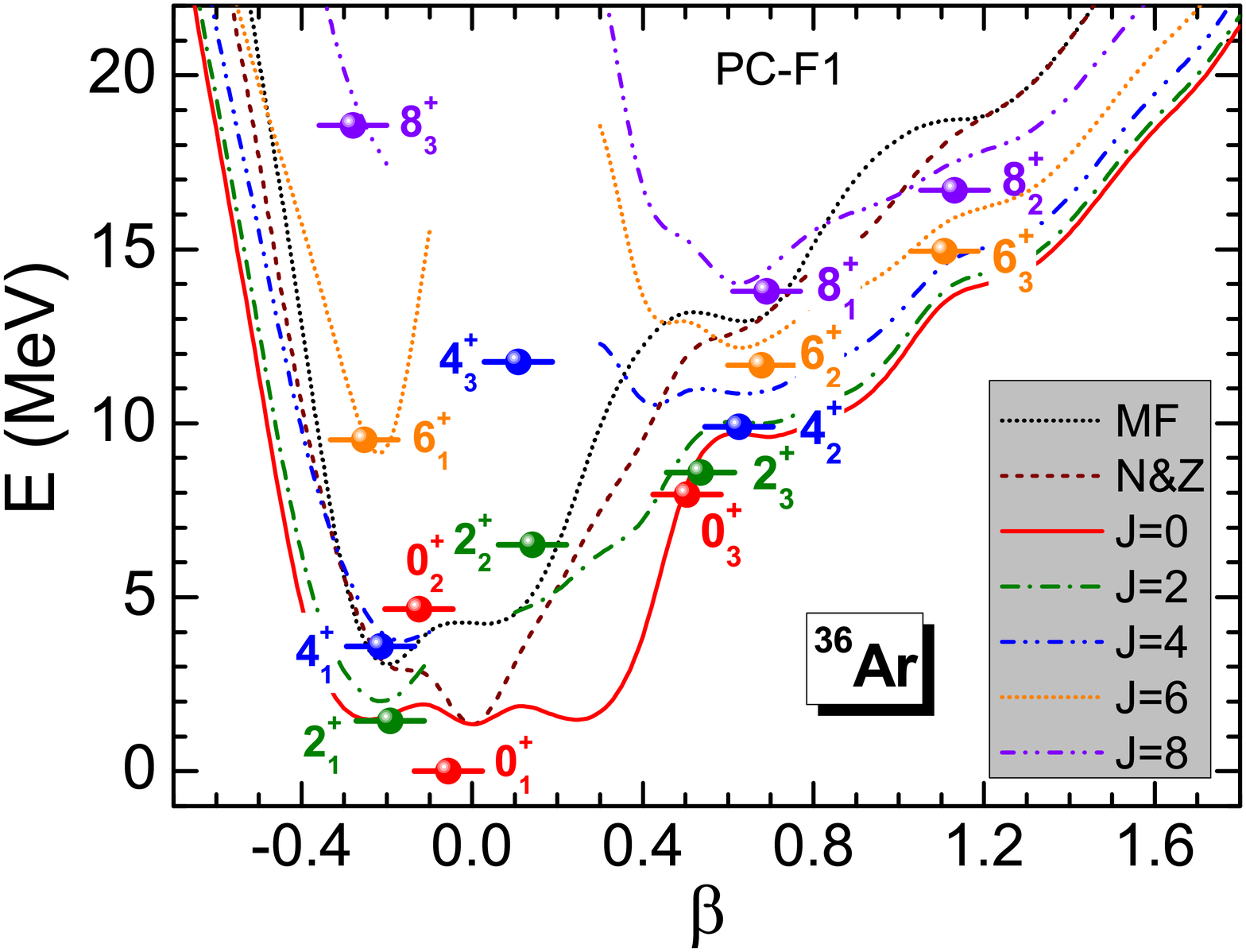}
 \caption{
  (Color online) Total energy (normalized to the $0_1^+$ state) for the mean-field states (MF), for
  the particle number projected states ($N\&Z$), and for the particle number and angular momentum projected
  states (with angular momentum $J$ = 0, 2, 4, 6, and 8) for $^{36}$Ar as a function of intrinsic mass quadrupole
  deformation. The solid bullets and the horizontal bars indicate the lowest GCM solutions which are
  plotted at their average deformation.
  }
 \label{fig:PECs_Ar36}
 \end{figure}

 The beyond-mean-field studies of the ND and SD states in $^{36}$Ar have been performed by several groups~\cite{Bender03,Rodriguez04,Niksic06}. Therefore, here we discuss very briefly our results for $^{36}$Ar, with an emphasis on the difference among the results of different models. Before spelling out our results, we note that our results for $^{36}$Ar might be somewhat different from those in Ref.~\cite{Niksic06} because of the different numerical details, such as the way to generate mean-field reference states and the treatment of particle number projection.

 Figure~\ref{fig:PECs_Ar36} displays the comparison of the energy curves for both mean-field and quantum-number projected states with $J$ = 0, 2, 4, 6, and 8. It shows that the energy gained from symmetry restoration changes significantly the topography of the energy curve. The energy curve of $J=0$ becomes rather flat around the spherical shape in the region $-0.3\leqslant\beta\leqslant0.3$. Moreover, the deformation of the SD state is shifted to $\beta=0.70$, compared to the mean-field value $\beta=0.64$. The discrete states from the GCM calculation, which are placed at their averaged quadrupole deformation $\bar\beta=\sum_{\beta} \vert g^J_\alpha (\beta) \vert^2 \beta$ with $g^J_\alpha (\beta)\equiv \sum_{\beta'}\left[{\cal N}^J(\beta,\beta')\right]^{1/2} f^{J}_{\alpha}(\beta')$, form one weakly (normal) deformed band and a well-deformed  rotational band.


 \begin{figure}[tbp]
 \centering
 \includegraphics[clip=,width=8.5cm]{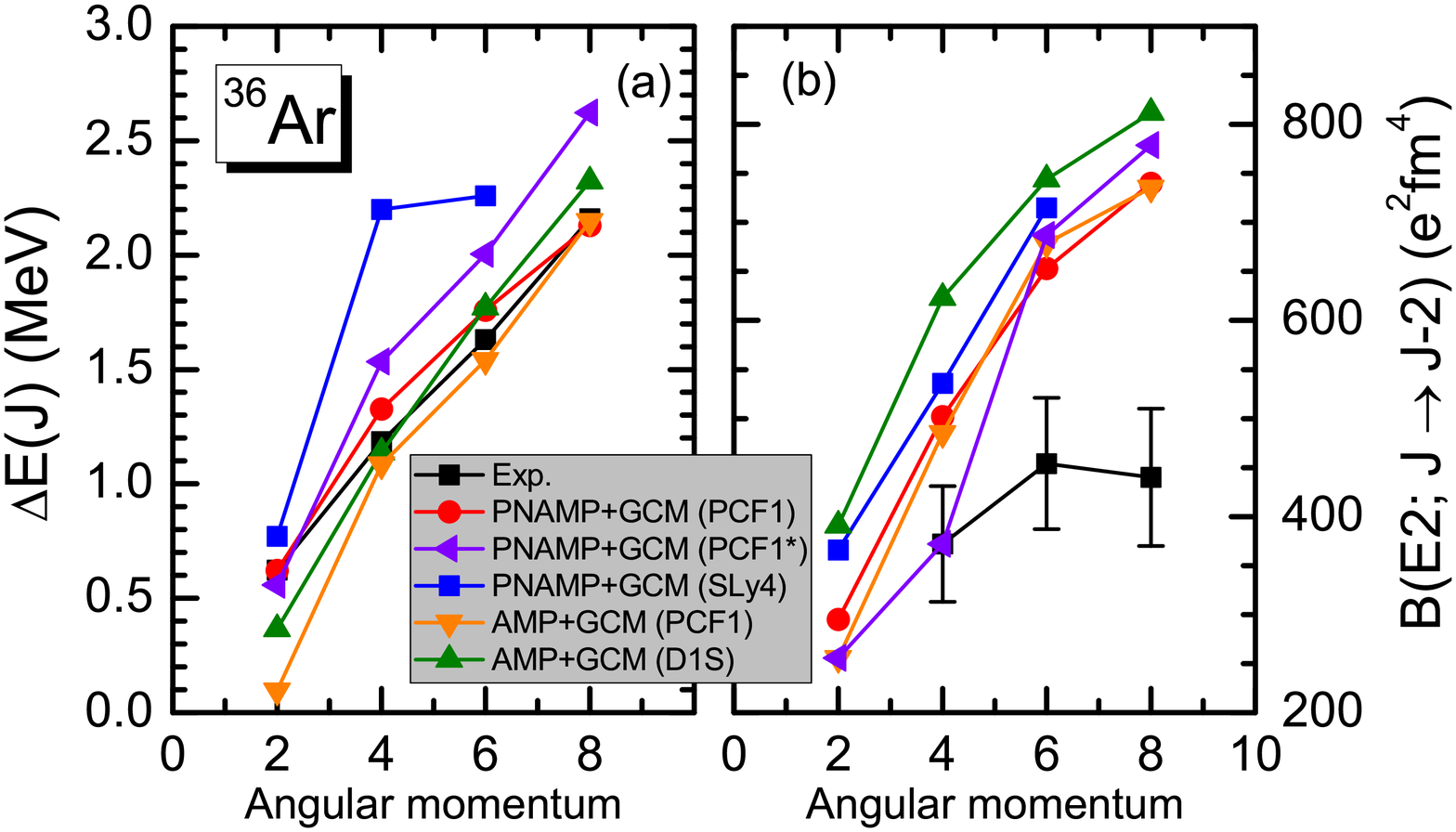}
 \caption{
  (Color online) (a) The energy difference $\Delta E(J)= E(J) - E(J-2)$ and (b) the reduced electric quadrupole transition strengtha $B(E2; J \rightarrow J-2)$ for the superdeformed (SD) states of $^{36}$Ar as a function of angular momentum. The results obtained by the PNAMP+GCM with Skyrme SLy6 interaction~\cite{Bender03}, the AMP+GCM with Gogny D1S interaction~\cite{Rodriguez04}, the PNAMP+GCM based on the RMF+LNBCS states with the re-adjusted PC-F1$^\ast$ interaction~\cite{Niksic06}, and the AMP+GCM with the PC-F1 interaction~\cite{Niksic06}, respectively, are plotted for comparison. Experimental data are taken from Refs.~\cite{Svensson01,NNDC}.
  }
 \label{fig:SDband}
 \end{figure}

 Figure~\ref{fig:SDband} shows the energy difference $\Delta E(J)= E(J) - E(J-2)$ and $B(E2)$ value as a function of angular momentum of the SD band in $^{36}$Ar. The excitation energy of the bandhead of the SD band is predicted to be around 8.0 MeV, compared with the value 5.9 MeV by the PNAMP+GCM based on the Skyrme SLy6 interaction~\cite{Bender03}, 7.5 MeV by the AMP+GCM with Gogny D1S interaction~\cite{Rodriguez04}, 9.2 MeV by the PNAMP+GCM with the re-adjusted PC-F1$^\ast$ interaction~\cite{Niksic06}, and 9.4 MeV by the AMP+GCM with the PC-F1 interaction~\cite{Niksic06}, respectively. However, the experimental value 4.3 MeV~\cite{Svensson01} is much smaller than all the predictions. The consideration of triaxiality and the effect of time-reversal symmetry breaking in the reference state may improve this description. Moreover, we note that the energy difference $\Delta E(J)$ between the SD states is well reproduced. However, the  $B(E2)$ values from all the model calculation are increasingly overestimated with angular momentum. It indicates again the possible increasing important role of the effect of time-reversal symmetry breaking with the angular momentum. However, this study is beyond the crurrent work.

 \begin{figure}[tbp]
 \centering
 \includegraphics[clip=,width=8cm]{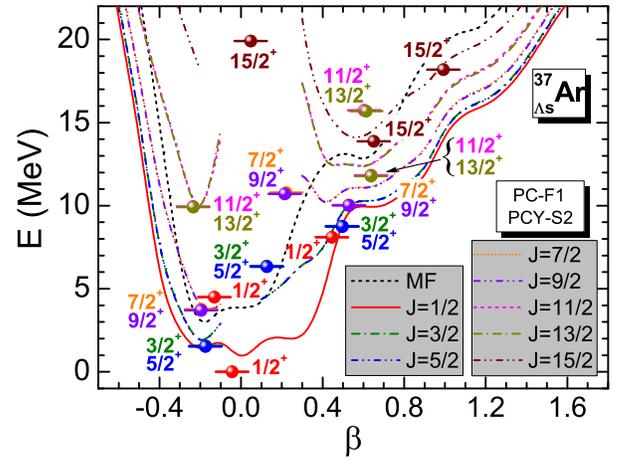}\vspace{-0cm}
 \caption{
  (Color online) The same as Fig.~\ref{fig:PECs_Ar36}, but for $^{37}_{\Lambda_s}$Ar.
  The parameter sets PC-F1 and PCY-S2 are adopted for the $NN$ and $N\Lambda$ interactions, respectively.}
 \label{fig:PECs_Ar37}
 \end{figure}

 The beyond-mean-field effect on hypernuclear states is investigated by taking the PC-F1 ($NN$) and PCY-S2 ($N\Lambda$) interactions and putting the $\Lambda$ in the lowest energy state. Fig.~\ref{fig:PECs_Ar37} displays the same quantities as those in Fig.~\ref{fig:PECs_Ar36} but for the beyond-mean-field calculation of $^{37}_{\Lambda_s}$Ar based on the PC-F1 ($NN$) + PCY-S2 ($N\Lambda$) interaction. We find that the deformation of the ND minimum of the energy curve $J=1/2$ of $^{37}_{\Lambda_s}$Ar is $\beta = - 0.20$, smaller than that ($\beta = - 0.25$) of $J=0$ energy curve in Fig.~\ref{fig:PECs_Ar36}. The deformation of the SD minimum is shifted from $\beta = 0.70$ (for $^{36}$Ar) to $\beta = 0.64$ (for $^{37}_{\Lambda_s}$Ar). After performing the configuration mixing calculation, we obtain the discrete hypernuclear states $J^+$ ($J=J_c\pm 1/2$) which are almost two-fold degenerate with the excitation energies close to those of the core states $J^+_c$. The similar phenomenon  has also been found in $^{21}_{\Lambda}$Ne~\cite{Mei16}. The $S_\Lambda$ for the ND $1/2^+$ state is 19.21 MeV and $S_\Lambda = 19.06$ MeV for the SD $1/2^+$ state. These values should be compared to the mean-field results of 19.28 MeV and 19.23 MeV, respectively. In other words, the beyond-mean-field effect decreases the $\Lambda_s$ binding energy of the SD state by 0.17 MeV, while it is nearly negligible for the ND state.

 \begin{figure}[tbp]
 \centering
 \includegraphics[clip=,width=8.5cm]{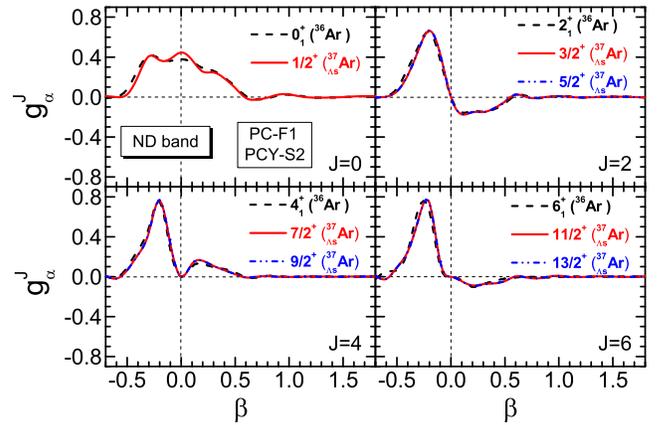}\vspace{-0cm}
 \caption{
  (Color online) Collective wavefunctions of the normal deformed states in $^{36}$Ar (dashed line) and $^{37}_{\Lambda_s}$Ar (solid line).
  The parameter sets PC-F1 and PCY-S2 are adopted for the $NN$ and $N\Lambda$ interactions, respectively.
  }
 \label{fig:NDwavfun}
 \end{figure}

 \begin{figure}[tbp]
 \centering
 \includegraphics[clip=,width=8.5cm]{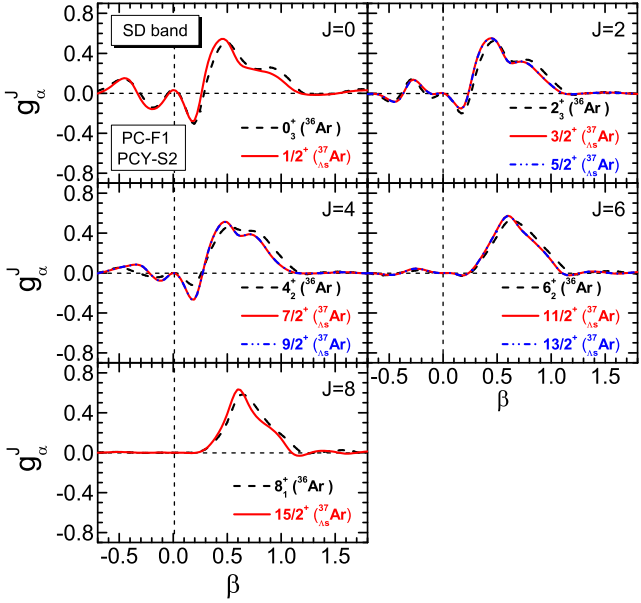}\vspace{-0cm}
 \caption{  (Color online) The same as Fig.~\ref{fig:NDwavfun}, but for the superdeformed (SD) states.}
 \label{fig:SDwavfun}
 \end{figure}

 The collective wavefunctions $g_\alpha^J$ for the ND and SD states in both $^{36}$Ar and $^{37}_{\Lambda_s}$Ar are plotted in Fig.~\ref{fig:NDwavfun} and Fig.~\ref{fig:SDwavfun}, respectively. It is shown that the wavefunctions for the two-fold degenerate states with $J=J_c\pm 1/2$ are almost on top of each other. Compared with those of $^{36}$Ar, the collective wavefunctions of hypernuclear states in $^{37}_{\Lambda_s}$Ar are slightly shifted inward to spherical shape. It is consistent with previous studies~\cite{Win08,Win11,Isaka12,Yao11,Mei14,Xue15,Mei15,Mei16} about the impurity effect of $\Lambda_s$ which reduces the quadrupole collectivity of atomic nuclei.

 \begin{figure}[tbp]
 \centering
 \includegraphics[clip=,width=8.5cm]{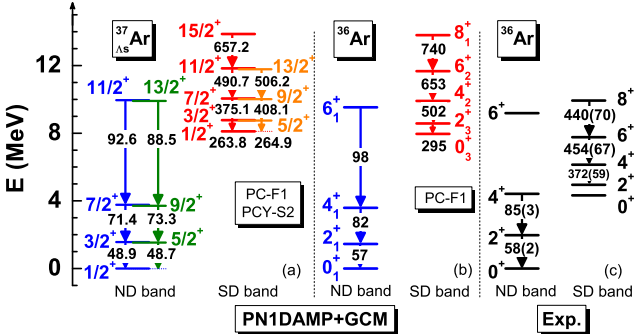}
 \caption{
  (Color online) The energy spectra of $^{37}_{\Lambda_s}$Ar (a) and $^{36}$Ar (b), in comparison with data (c)~\cite{Svensson01,NNDC}. The reduced electric quadrupole transition strengths $B(E2)$ (in units of $e^2$ fm$^4$) are provided on the arrows. The parameter sets PC-F1 and PCY-S2 are adopted for the $NN$ and $N\Lambda$ interactions, respectively. }
 \label{fig:spectrum}
 \end{figure}

 Figure~\ref{fig:spectrum} displays the energy spectra of $^{36}$Ar and $^{37}_{\Lambda_s}$Ar, in comparison with available data of $^{36}$Ar. We note that the energies and $E2$ transition strengths for the ND states are reproduced rather well. Again the SD states are systematically overestimated. Compared to $^{36}$Ar, the $E2$ transition strength between the ND $3/2^+, 1/2^+$ states in $^{37}_{\Lambda_s}$Ar is 48.9 $e^2$ fm$^4$, smaller than the $B(E2; 2^+ \rightarrow 0^+$) in $^{36}$Ar by $14.5\%$, while the $E2$ transition strength between the SD $3/2^+, 1/2^+$ states is reduced by $10.7\%$. Moreover, the excitation energy of  the ND and SD $3/2^+$ states in $^{37}_{\Lambda_s}$Ar is found by $7.7\%$ and $2.2\%$ larger than those of the $2^+_1$ state in $^{36}$Ar,  respectively. It hints that the $\Lambda_s$ hyperon impurity effect on the energy spectra is more pronounced for the ND state than for the SD state in $^{37}_{\Lambda_s}$Ar.

%
\section{summary}\label{Sec.V}
%

 We have presented both mean-field and beyond-mean-field studies for the hyperon impurity effect in $^{37}_\Lambda$Ar with the coexistence of ND and SD shapes in the case of the $\Lambda$ is put in the lowest one of the states which correspond to the $s, p$, or $d$ state in the spherical limit, respectively. In the mean-field calculation, four sets of relativistic point-coupling $N\Lambda$ interactions PCY-S1, PCY-S2, PCY-S3, and PCY-S4 have been adopted to examine the parameter-dependence of the results. To scrutinize the beyond-mean-field effect, we have carried out a quantum number (particle number and angular momentum) projected generator coordinate method calculation for $^{37}_{\Lambda s}$Ar.

 Our results indicate that after taking the hyperon impurity effect into account, the SD states persist in $^{37}_{\Lambda}$Ar for all the four $N\Lambda$ effective interactions and the $\Lambda_s$ decreases the quadrupole collectivity of ND states to a greater extent than that of SD states. Moreover, the beyond-mean-field effect decreases the $\Lambda_s$ binding energy in the SD state by 0.17 MeV, while its effect on that of the ND state is negligible. The predicted larger $\Lambda_s$ separation energy in the SD state by relativistic models is not necessary attributed to the ring-shaped clustering structure of nucleons in hypernuclei. The distribution of the hyperon, which depends on the details of the $N\Lambda$ interaction, may play a more important role. The $\Lambda_p$ and $\Lambda_d$ binding energies of SD states are always larger than those in the ND states. Finally, we point out that the SD states of hypernuclei might be difficult to be produced in current experimental facilities, the conclusions derived from this study are helpful to understand the hyperon impurity effect on nuclear matter and atomic nuclei in a comprehensive way.


 \begin{acknowledgements}
  The authors express their deep gratitude to Professor K. Hagino for constructive suggestions and remarks.
  This work was supported by the National Natural Science Foundation of China under
  Grant Nos. 11275160, 11575148, 11475140, 11305134.
 \end{acknowledgements}

 

\end{document}